\documentclass[%
 reprint,
 amsmath,amssymb,
 aps,
]{revtex4-2}

\usepackage{graphicx}% Include figure files
\usepackage{dcolumn}% Align table columns on decimal point
\usepackage{bm}% bold math
\usepackage{enumerate}
\begin{document}

\preprint{APS/123-QED}

\title{Macroscopic dynamics of oscillator ensembles with communities, higher-order interactions, and phase lags}% Force line breaks with \\

\author{Sabina Adhikari}
 \email{sabina.adhikari@colorado.edu}
\author{Juan G. Restrepo}%
 \email{juanga@colorado.edu}
\affiliation{%
 Department of Applied Mathematics, University of Colorado Boulder, Boulder, Colorado, 80309
}%

\author{Per Sebastian Skardal}
 \email{persebastian.skardal@trincoll.edu}
\affiliation{
 Department of Mathematics, Trinity College, Hartford, Connecticut, 06106% with \\
}%

\date{\today}% It is always \today, today,

\begin{abstract} 
We study the effects of phase-frustrated, higher-order interactions in a system of coupled phase oscillators with two communities. We use dimensionality reduction techniques to derive a low-dimensional system of ODEs to describe the macroscopic behavior of the system. By analyzing this system we show that, in addition to the fixed point solutions present in a system of oscillators with higher order interactions and community structure only, the system also exhibits oscillatory or chaotic synchronization behavior for some phase lag values. Moreover, some phase lag values give rise to multistability of solutions, where both fixed point solutions and oscillatory or chaotic behavior of the order parameters can be observed, depending on initial conditions. 
\end{abstract}

\maketitle

%\tableofcontents

\section{\label{Introduction}Introduction}

Synchronization is a collective behavior emergent in many complex systems, including applications in physics \cite{zhu2015synchronization}, biology \cite{petri2014homological, kitzbichler2009broadband}, neuroscience \cite{neurons1, neuroscience}, and engineering \cite{strogatz2005crowd, fujino1993synchronization, Powergrid}. Examples of systems exhibiting synchronization include flashing fireflies \cite{sarfati2020spatio, Fireflies}, neuronal networks \cite{neurons1}, power grids \cite{Powergrid, Powergrid1}, circadian rhythms \cite{circadian}, and Josephson junctions \cite{PhysRevLett.76.404}. The Kuramoto model \cite{kuramoto1975self} and its generalization, the Kuramoto-Sakaguchi model \cite{guchi}, have been widely used to understand the dynamics of weakly-coupled phase oscillators. While these models describe synchronization mediated via pairwise interactions between oscillators, they do not account for higher-order interactions: nonlinear, simultaneous interactions between multiple oscillators. Recently, much effort has been dedicated to understand the effects of these higher-order interactions on synchronization due to their relevance in neuroscience \cite{petri2014homological, Brain-simplex}, ecology \cite{ecology-higher-order}, and physics \cite{higherorder-physics, PhysRevE.100.012211}. Higher-order interactions in coupled oscillator systems can result in interesting dynamics, such as bistability, explosive transitions, and hysteresis \cite{PhysRevLett.122.248301, skardal-arenas1, Sync-hypergraphs, PhysRevLett.124.218301}.

The Kuramoto-Sakaguchi model \cite{guchi}, which captures phase-frustrated coupling in coupled oscillator systems, is given by
\begin{eqnarray}
    \frac{d \theta_i}{dt} = \omega_i + \frac{K}{N} \sum_{j=1}^N \sin{(\theta_j - \theta_i - \gamma)}, \label{eq:kuramoto-sakaguchi1}
\end{eqnarray}
where $\theta_i$ and $\omega_i$ are, respectively, the phase and intrinsic frequency of oscillator $i$, $K$ is the coupling strength, and the phase lag parameter $\gamma$ is assumed to be in the interval $[0, \pi/2]$. Phase-frustrated coupling in pairwise-interacting oscillator systems results in rich phenomena such as nonuniversal transitions to synchrony \cite{PhysRevLett.109.164101}, chimera states \cite{chimera1, PhysRevLett.93.174102, PhysRevE.93.012218} and chaos \cite{Chaos-phaselags}. In Ref.~\cite{Chaos-phaselags}, the authors show that heterogeneous phase lags can result in chaos in a network of coupled oscillators with community structure. In Ref.~\cite{PhysRevE.108.034208}, the authors find that phase lags can either induce or suppress explosive transitions between incoherent and synchronized states in coupled phase oscillator systems with higher-order interactions. Similarly, in Ref.~\cite{das2025effectphaselagsynchronizationadaptive} the authors show that phase lags can either inhibit or promote tiered and explosive synchronization transitions in adaptive multilayer networks with higher-order interactions. Despite the prevalence of phase lags and community structure in many systems, the combined effects of phase lags in oscillator systems with higher-order interactions and community structure remain unexplored. Hence, following \cite{PhysRevE.100.012211, Multistability-communities, PhysRevLett.122.248301,Sync-hypergraphs}, we study a generalization of the Kuramoto-Sakaguchi model with higher-order interactions and community structure. Using the Ott-Antonsen ansatz \cite{Ott-Antonsen}, we obtain a system of low-dimensional equations to describe the macroscopic order parameters. We focus on oscillator networks with pairwise and triadic coupling, two communities, and equal phase lags between a pair and a triplet of oscillators. Using the low dimensional equations and numerical simulations, we show that the coupled oscillator systems can exhibit a variety of dynamics including stable synchronized states, periodic synchronization, or chaotic synchronization behavior depending on the values of the phase lags.

This paper is organized as follows. In Sec.~\ref{Model}, we present a generalization of Kuramoto-Sakaguchi model. In Sec.~\ref{Dimensionality Reduction}, we use the Ott-Antonsen ansatz to derive low dimensional equations for the system's order parameters. In Sec.~\ref{Two Communities}, we focus on two communities and define the notation for this case. In Sec.~\ref{negativecoupling} and Sec.~\ref{positivecoupling}, we present results for two different cases: negative inter-community coupling strength and positive inter-community coupling strength, respectively. In Sec.~\ref{Discussion}, we discuss our results and present a conclusion.

\section{\label{Theory} Model and Theory}

\subsection{Model}\label{Model}

We consider a population of $N$ phase oscillators divided into $C$ communities. We assume that each community $\sigma$ has $N_{\sigma}$ oscillators, where $\sigma = 1, 2, \cdots, C$ and $\sum_{\sigma=1}^C N_{\sigma} = N$. Here we consider synchronization mediated by dyadic and triadic interactions with phase lags. Following Refs.~\cite{PhysRevLett.122.248301,skardal-arenas1,Sync-hypergraphs,Multistability-communities}, we assume that the oscillator population evolves according to

\begin{widetext}
\begin{eqnarray}
    \frac{d \theta_i^{\sigma}}{d t} = \omega_i^{\sigma} + \sum_{\sigma' = 1}^C \frac{K_{1}^{\sigma \sigma'}}{N_{\sigma'}} \sum_{j=1}^{N_{\sigma'}} \sin{(\theta_j^{\sigma'} - \theta_i^{\sigma} - \gamma_{\sigma \sigma'})} + \sum_{\sigma' = 1}^C \sum_{\sigma'' = 1}^C \frac{K_{2}^{\sigma \sigma' \sigma''}}{N_{\sigma'} N_{\sigma''}} \sum_{j=1}^{N_{\sigma'}} \sum_{l=1}^{N_{\sigma''}} \sin{(2\theta_j^{\sigma'} -\theta_l^{\sigma''} - \theta_i^{\sigma} - \gamma_{\sigma \sigma' \sigma''})}, \label{Kuramoto-equation}
\end{eqnarray}
\end{widetext}
where $\theta_i^{\sigma}$ and $\omega_i^{\sigma}$ denote, respectively, the phase and natural frequency of an oscillator $i$ in community $\sigma$. The quantities $K_1^{\sigma \sigma'}$ and $\gamma_{\sigma \sigma'}$ denote the dyadic coupling strength and phase lag, respectively, between a pair of oscillators in communities $\sigma$ and $\sigma'$. Similarly, $K_2^{\sigma\sigma'\sigma''}$ and $\gamma_{\sigma\sigma'\sigma''}$ denote the triadic coupling strength and phase lag between a triplet of oscillators in communities $\sigma$, $\sigma'$, and $\sigma''$. For simplicity, we will consider only the case where $ \gamma_{\sigma \sigma'} = \gamma_{\sigma\sigma'\sigma''} = \gamma$, for $\gamma \in [0, \pi/2]$. We note that higher-order interactions of the form (\ref{Kuramoto-equation}) arise from a higher-order expansion of the complex Ginzburg-Landau equation (Refs.~\cite{PhysRevE.100.012211,complex-ginzburg2}) when carrying out a phase reduction procedure. The dynamics of Eq.~(\ref{Kuramoto-equation}) without higher-order interactions has been previously studied for two communities both in the case of identical phase lags \cite{guchi, PhysRevLett.101.084103, PhysRevE.93.012218} and with distinct phase lags and coupling strengths between communities \cite{complex-ginzburg2,Chaos-phaselags}. In this paper, we consider a population of oscillators with two communities, higher-order interactions, and identical phase lags between interacting groups of oscillators.

We define the community-wise order parameter
\begin{eqnarray}
z^{\sigma}_q &=& r^{\sigma}_q e^{i \psi^{\sigma}_q} = \frac{1}{N_{\sigma}}\sum_{j=1}^{N_\sigma} e^{q i \theta_j^{\sigma}} \label{order parameter eq},   
\end{eqnarray}
where $q=1$ corresponds to the classical Kuramoto order parameter and $q=2$ measures cluster synchrony \cite{PhysRevE.84.036208}. Using these community-wise order parameters, Eq.~(\ref{Kuramoto-equation}) can be written as
\begin{eqnarray}
        \dot{\theta^{\sigma}_i} &= \omega_i^{\sigma} + \frac{1}{2i}\left(H e^{-i \theta^\sigma_i} - H^* e^{i \theta^\sigma_i}\right), \label{Kuramoto-eqn2}
\end{eqnarray}
where 
\begin{eqnarray}
    H = e^{-i \gamma} \left(\sum_{\sigma'=1}^C K_1^{\sigma \sigma'} z_1^{\sigma'} + \sum_{\sigma'=1}^{C} \sum_{\sigma'' = 1}^{C} K_2^{\sigma \sigma' \sigma''} z_2^{\sigma'} (z_1^{\sigma''})^*\right) \nonumber
    \\
\end{eqnarray} 
and * indicates complex conjugate.
\subsection{Dimensionality Reduction \label{Dimensionality Reduction}}

In this section we use the Ott-Antonsen ansatz \cite{Ott-Antonsen} to derive low-dimensional equations for the macroscopic order parameters. Moving to the continuum limit $N \to \infty$, we define $f^{\sigma}(\theta,\omega,t)$ as the density of oscillators with phase $\theta$ and natural frequency $\omega$ at time $t$ in community $\sigma$. The density function $f^{\sigma}$ satisfies the continuity equation due to conservation of oscillators, 
\begin{eqnarray}
    \frac{df^{\sigma}}{dt} + \frac{d}{d\theta}(f^{\sigma} \dot{\theta^{\sigma}}) &=& 0. \label{continuty eqn}
\end{eqnarray}
We then write $f^{\sigma}$ as a Fourier series,
\begin{eqnarray}
    f^{\sigma} &=& \frac{g^{\sigma}(\omega^{\sigma})}{2\pi} \left[ 1 + \sum_{n=1}^{\infty} f_n^{\sigma} e^{i n \theta} + c.c. \right], \label{fourier-series}
\end{eqnarray}
where $g^{\sigma}(\omega^{\sigma})$ is the natural frequency distribution of oscillators in community $\sigma$ and $c.c.$ denotes complex conjugate of the preceding term. We use the Ott-Antonsen ansatz for the Fourier modes, i.e., $f_n^{\sigma}(\omega,t) = (b^{\sigma}(\omega,t))^n$. Substituting this ansatz into Eq.~(\ref{continuty eqn}), we find
\begin{eqnarray}
    \frac{d b^{\sigma}}{dt} + i \omega^{\sigma} b^{\sigma} - \frac{1}{2}[H - H^*(b^{\sigma})^2] &=& 0. \label{fouriermodeeqn}
\end{eqnarray}
Note that the order parameter $z^{\sigma}_q$ can be written as
\begin{eqnarray}
    z^{\sigma}_q = \int_{-\infty}^{\infty} \int_{0}^{2\pi} f^{\sigma}(\theta,\omega,t) e^{q i \theta} d\theta d\omega. \label{orderint}
\end{eqnarray}
Integrating Eq.~(\ref{orderint}) with respect to $\theta$, we can write $z^{\sigma}_q$ in terms of the Fourier modes
\begin{eqnarray}
    z^{\sigma}_{q} = \int_{-\infty}^{\infty} (b^{\sigma*}(\omega,t))^q g^{\sigma}(\omega) d\omega. \label{orderint1}
\end{eqnarray}

Assuming a Lorentzian distribution of frequencies, $g(\omega) = {\Delta^{\sigma}}/({\pi [(\Delta^{\sigma})^2 + (\omega^{\sigma} - \omega^{\sigma}_0)^2}])$, we use contour integration to solve Eq.~(\ref{orderint1}) and get
\begin{eqnarray}
    z_1^{\sigma}(t) = b^{\sigma*}(\omega_0^{\sigma} - i \Delta^{\sigma}, t), \label{ordepara-beqn}
\end{eqnarray}
and 
\begin{eqnarray}
    z_2^{\sigma} = [z_1^{\sigma}(t)]^2.
\end{eqnarray}
Substituting Eq.~(\ref{ordepara-beqn}) into Eq.~(\ref{fouriermodeeqn}), taking the complex conjugate and, writing $z_1^{\sigma}$ simply as $z^{\sigma}$, we get a system of ODEs for the community-wise order parameters,
\begin{eqnarray}
    \dot{z}^{\sigma} &=& i \omega_0^{\sigma}z^{\sigma} - \Delta^{\sigma}z^{\sigma} + \frac{1}{2}\left\{H - H^* (z^{\sigma})^2\right\}. \label{z-ode}
\end{eqnarray}
As $z^{\sigma} = r^{\sigma}e^{i \psi^{\sigma}}$, Eq.~(\ref{z-ode}) can be written as
\begin{widetext}
\begin{eqnarray}
    \dot{r}^{\sigma} e^{i \psi^{\sigma}} + i r^{\sigma} e^{i \psi^{\sigma}} \dot{\psi}^{\sigma} &=& i \omega_0^{\sigma} r^{\sigma} e^{i \psi^{\sigma}} - \Delta^{\sigma} r^{\sigma} e^{i \psi^{\sigma}} + \frac{1}{2} \left[ \sum_{\sigma'= 1}^CK_{1}^{\sigma \sigma'} r^{\sigma'} \left(e^{i \psi^{\sigma'}} e^{-i \gamma} - (r^{\sigma})^2 e^{-i \psi^{\sigma'}} e^{i \gamma} e^{2 i \psi^{\sigma}}\right)\right] \label{polareqn}\\
    & + & \frac{1}{2}\left[\sum_{\sigma'=1}^C \sum_{\sigma''=1}^C K_2^{\sigma \sigma' \sigma''}(r^{\sigma'})^2 r^{\sigma''} \left(e^{2i \psi^{\sigma'}} e^{-i \psi^{\sigma''}} e^{-i \gamma} - (r^{\sigma})^2 e^{-2i \psi^{\sigma'}} e^{i \psi^{\sigma''}} e^{i \gamma} e^{2i \psi^{\sigma}} \right) \right]. \nonumber
\end{eqnarray}
\end{widetext}
We separate the real and the imaginary parts of Eq.~(\ref{polareqn}) to get the equations for $r^{\sigma}$ and $\psi^{\sigma}$,
\begin{widetext}
\begin{eqnarray}
    \dot{r}^{\sigma} &=& -\Delta^{\sigma} r^{\sigma} + \frac{[1 - (r^{\sigma})^2]}{2} \sum_{\sigma'=1}^C K_1^{\sigma \sigma'} r^{\sigma'} \cos{(\psi^{\sigma'} - \psi^{\sigma} - \gamma)} \nonumber\\
    &+& \frac{[1 - (r^{\sigma})^2]}{2}\left[\sum_{\sigma'=1}^C \sum_{\sigma''=1}^C K_2^{\sigma \sigma' \sigma''} (r^{\sigma'})^2 r^{\sigma''} \cos{(\psi^{2\sigma'} - \psi^{\sigma''}- \psi^{\sigma} -\gamma)}\right], \label{r-ode}
\end{eqnarray}    
\end{widetext}
and
\begin{widetext}
    \begin{eqnarray}
        \dot{\psi^{\sigma}} &=& \omega_0^{\sigma} + \frac{[1 + (r^{\sigma})^2]}{2r^{\sigma}} \sum_{\sigma'=1}^C K_1^{\sigma \sigma'} r^{\sigma'} \sin{(\psi^{\sigma'} - \psi^{\sigma} -\gamma)} \nonumber \\
        &+& \frac{[1 + (r^{\sigma})^2]}{2r^{\sigma}}\left[\sum_{\sigma'=1}^C \sum_{\sigma''=1}^C K_2^{\sigma \sigma' \sigma''} (r^{\sigma'})^2 r^{\sigma''} \sin{(\psi^{2\sigma'} - \psi^{\sigma''}- \psi^{\sigma} - \gamma})\right]. \label{psi-ode}
    \end{eqnarray}
\end{widetext}
We will now examine Eqs.~(\ref{r-ode}) and (\ref{psi-ode}) specifically for an ensemble of oscillators divided into two communities.

\subsection{Two Communities \label{Two Communities}}
Before delving into the detailed analysis of Eqs.~(\ref{r-ode}) and (\ref{psi-ode}) for the case of two communities, we define the dyadic and triadic coupling strengths, $K_1^{\sigma \sigma'}$ and $K_2^{\sigma \sigma' \sigma''}$, respectively. Following \cite{Multistability-communities}, we introduce a parameter $\alpha \in (-1,1)$ and define the dyadic coupling strength as (assuming $\sigma \neq \sigma'$, $\sigma' \neq \sigma''$, and $\sigma \neq \sigma''$)
\begin{eqnarray}
    K_1^{\sigma \sigma} = K_1 \quad \text{and} \quad K_1^{\sigma \sigma'} = \alpha K_1, \label{dyadic-coupling}
\end{eqnarray}
and the triadic coupling strength as
\begin{eqnarray}
    K_2^{\sigma \sigma \sigma} = K_2 \quad \text{and} \quad K_2^{\sigma \sigma \sigma'} = K_2^{\sigma \sigma' \sigma} = \alpha K_2, \nonumber \\
    K_2^{\sigma \sigma' \sigma'} = \alpha^2 K_2 \quad \text{and} \quad K_2^{\sigma \sigma' \sigma''} = \alpha^3 K_2 \label{triadic-coupling}.
\end{eqnarray}

Defining $K_1^{\sigma \sigma'}$ as in Eq.~(\ref{dyadic-coupling}), we make the assumption that the coupling strength between a pair of oscillators belonging to the same community is stronger than the strength between a pair belonging to different communities. Similarly, defining $K_2^{\sigma \sigma' \sigma''}$ as in Eq.~(\ref{triadic-coupling}), we assume that the triadic coupling strength to an oscillator $i$ decreases with the fraction of the oscillators in the triad that belong to a different community as that of oscillator $i$. Note that $K_2^{\sigma \sigma' \sigma''}$ does not affect the dynamics of Eqs.~(\ref{r-ode}), (\ref{psi-ode}) as we only consider two communities here. Substituting Eqs.~(\ref{dyadic-coupling}) and (\ref{triadic-coupling}) into Eqs.~(\ref{r-ode}) and (\ref{psi-ode}), we get the following ODEs for ${r_1}$ and ${r_2}$,
\begin{widetext}
 \begin{eqnarray}
        \dot{r}_1 &=& -\Delta_1 r_1 + \frac{(1-r_1^2)}{2}\left[K_1 r_1 \cos{(\gamma)} + \alpha K_1 r_2 \cos{(\Phi - \gamma)} + K_2 r_1^3 \cos{(\gamma)} \right. \nonumber\\ 
        & & \left. + \alpha K_2 r_1^2 r_2 \cos{(\Phi + \gamma)}  + \alpha K_2 r_2^2 r_1 \cos{(2\Phi - \gamma)} + \alpha^2 K_2 r_2^3 \cos{(\Phi-\gamma)}\right] \label{r1-ode}, \\
        \dot{r}_2 &=& -\Delta_2 r_2 + \frac{(1-r_2^2)}{2} \left[\alpha K_1 r_1 \cos{(\Phi + \gamma)} + K_1 r_2 \cos{(\gamma)} +\alpha^2 K_2 r_1^3 \cos{(\Phi + \gamma)} \right.\nonumber \\
        && \left. + \alpha K_2 r_1^2 r_2 \cos{(2 \Phi + \gamma)}  + \alpha K_2 r_2^2 r_1 \cos{(\Phi - \gamma)} + K_2 r_2^3 \cos{(\gamma)}\right] \label{r2-ode}.
\end{eqnarray}   
\end{widetext}
We further reduce the dimension of our system by introducing a phase difference $\Phi = \psi^{2} - \psi^{1}$, yielding
\begin{widetext}
    \begin{eqnarray}
         \dot{\Phi} &=& \omega_0^{2} -\omega_{0}^1 + \frac{(1+r_2^2)}{2r_2} \left[K_1r_2\sin{(-\gamma)} + \alpha K_1 r_1 \sin{(-(\Phi +\gamma))} + K_2 r_2^3\sin{(-\gamma)} \right. \nonumber \\ 
        && \left. +\alpha K_2 r_1^2 r_ 2 \sin{(-(2\Phi+\gamma))} + \alpha K_2 r_1 r_2^2 \sin{(\Phi - \gamma)}+\alpha^2 K_2 r_1^3 \sin{(-(\Phi+\gamma))} \right] \nonumber \\ 
        && - \frac{(1+r_1^2)}{2r_1} \left[K_1r_1\sin{(-\gamma)} + \alpha K_1 r_2 \sin{(\Phi -\gamma)} + K_2 r_1^3\sin{(-\gamma)} \right. \nonumber \\
        && \left. +\alpha K_2 r_1^2 r_2 \sin{(-(\Phi+\gamma))} + \alpha K_2 r_1 r_2^2 \sin{(2\Phi - \gamma)}+\alpha^2 K_2 r_2^3 \sin{(\Phi-\gamma)}\right] \label{phi-ode}.
    \end{eqnarray}
\end{widetext}

In the forthcoming sections, we examine Eqs.~(\ref{r1-ode})-(\ref{phi-ode}) in detail for two different cases: $-1 < \alpha < 0$ and $0 < \alpha < 1$. For the rest of this work, we assume that oscillators in both communities have frequencies drawn from an identical Lorentzian distribution, as we are mainly interested in understanding the effects of higher-order interactions, community structure, and phase lags. Hence, we set $\Delta_1 = \Delta_2 = \Delta$ and $\delta \omega = \omega_0^2 - \omega_0^1 = 0$.

\section{Results}
%%%%%%%%%%%%%%%%
\begin{figure}[b]
    \centering    
    \includegraphics[width=\linewidth]{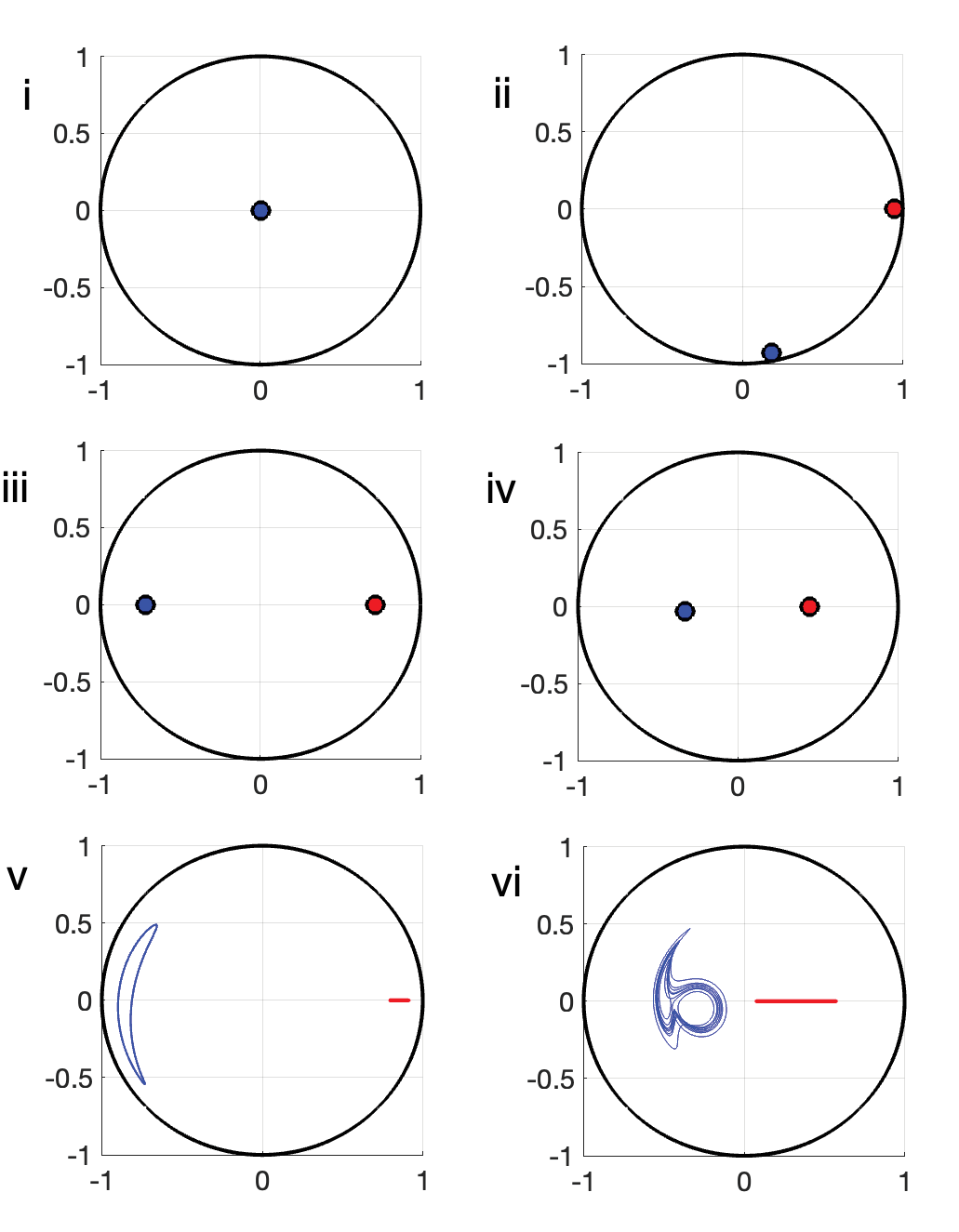}
    \caption[Six of the seven possible steady states for negative $\alpha$]{{\it Six possible steady states for negative $\alpha$.} Schematic illustration of the first six possible stable states in the two community case for negative $\alpha$: (i) incoherent state, (ii) skew-phase synchronized state, (iii) anti-phase synchronized state, (iv) almost anti-phase synchronized state, (v) periodic states, and (vi) chaotic states. Red and blue filled circles denote the steady states of the two communities. The blue and red curves in (v) and (vi) give periodic and chaotic trajectories.}
    \label{fig:steady-state-negative-alpha-complex-plane}
\end{figure}
%%%%%%%%%%%%%%%%
\subsection{Repulsive coupling, $-1 < \alpha < 0$}\label{negativecoupling}
Negative values of $\alpha$ imply repulsive coupling strength for oscillators in different communities, but cooperative coupling strength for oscillators within a community. For negative coupling the dynamics is very rich and a variety of states can arise, which we detail below.

\subsubsection{States observed and bifurcation diagram}

First, we note that for $\gamma = 0$ the two communities reach the steady states found in \cite{Multistability-communities}: an {\it incoherent} state with $r_1 = r_2 = 0$ and a {\it skew-phase synchronized} state with $r_1 = r_2 >0$ and $0 < \Phi < \pi/2$. For $\gamma > 0$, the communities exhibit seven different possible dynamics: 
\begin{enumerate}[(i)]
\item The {\it incoherent state} with $r_1 = r_2 = 0$.
\item The {\it skew-phase synchronized state} with $r_1 = r_2 > 0$ and $0 < \Phi < \pi/2$.
\item The {\it anti-phase synchronized state} with $r_1 = r_2 > 0$ and $\Phi = \pi$.
\item {The \textit{almost anti-phase state}} with constant $r_1$ and $r_2$, $r_1 > 0$, $r_2 > 0,  r_1 \neq r_2 \text{ and } \Phi \approx \pi$.
\item {\it Periodic states} with $r_1(t) = r_1(t + \tau)$ and $r_2(t) = r_2(t + \tau)$ for some $\tau > 0$. 
\item {\it Chaotic states} where $r_1$ and $r_2$ exhibit chaotic behavior.
\item {The \textit{asymmetric skew-phase state}} with $r_1 \approx r_2$ and $0 < \Phi < \pi/2$.

\end{enumerate}
%%%%%%%%%%%%%%%%%%%%%%%%%
\begin{figure}[t]   
 \includegraphics[width=0.49\linewidth]
 {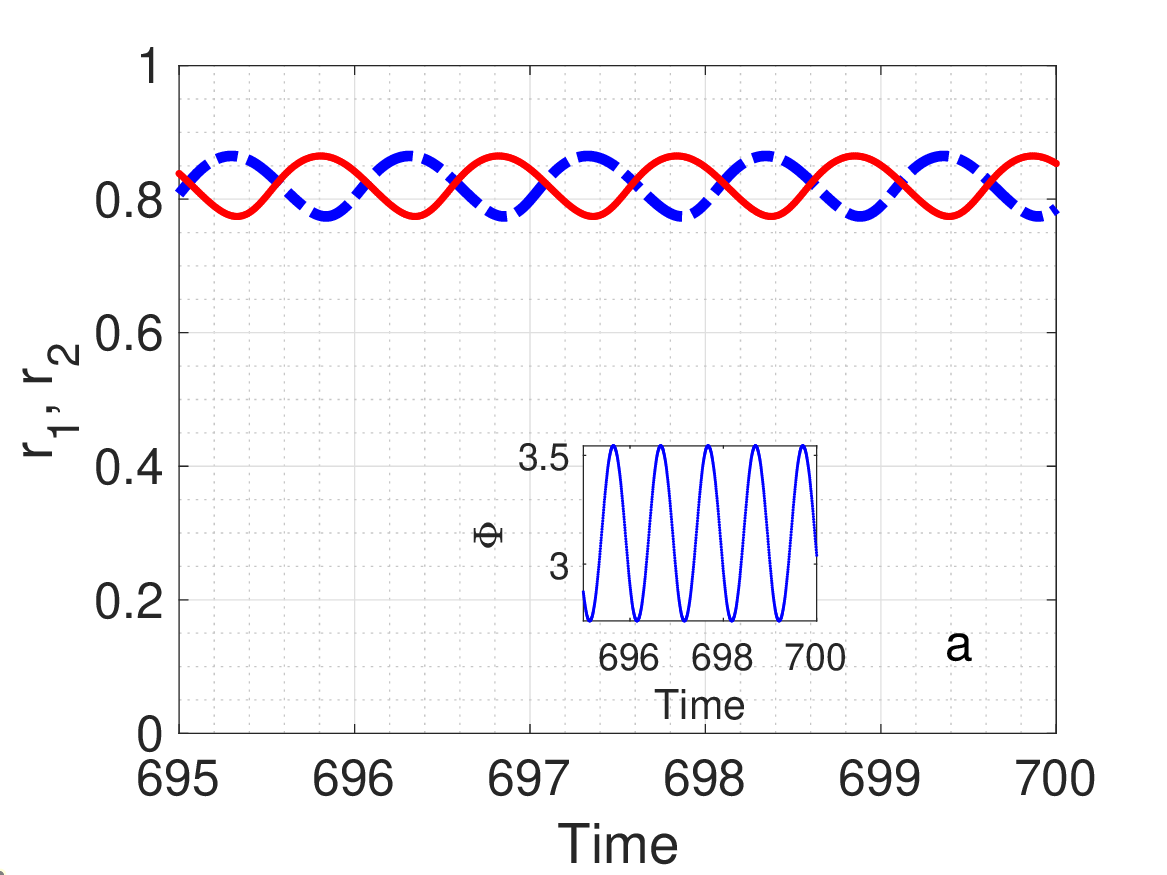}%{polar_timeseries_1.25_neg_alpha.eps}  
  \includegraphics[width=0.49\linewidth]
  {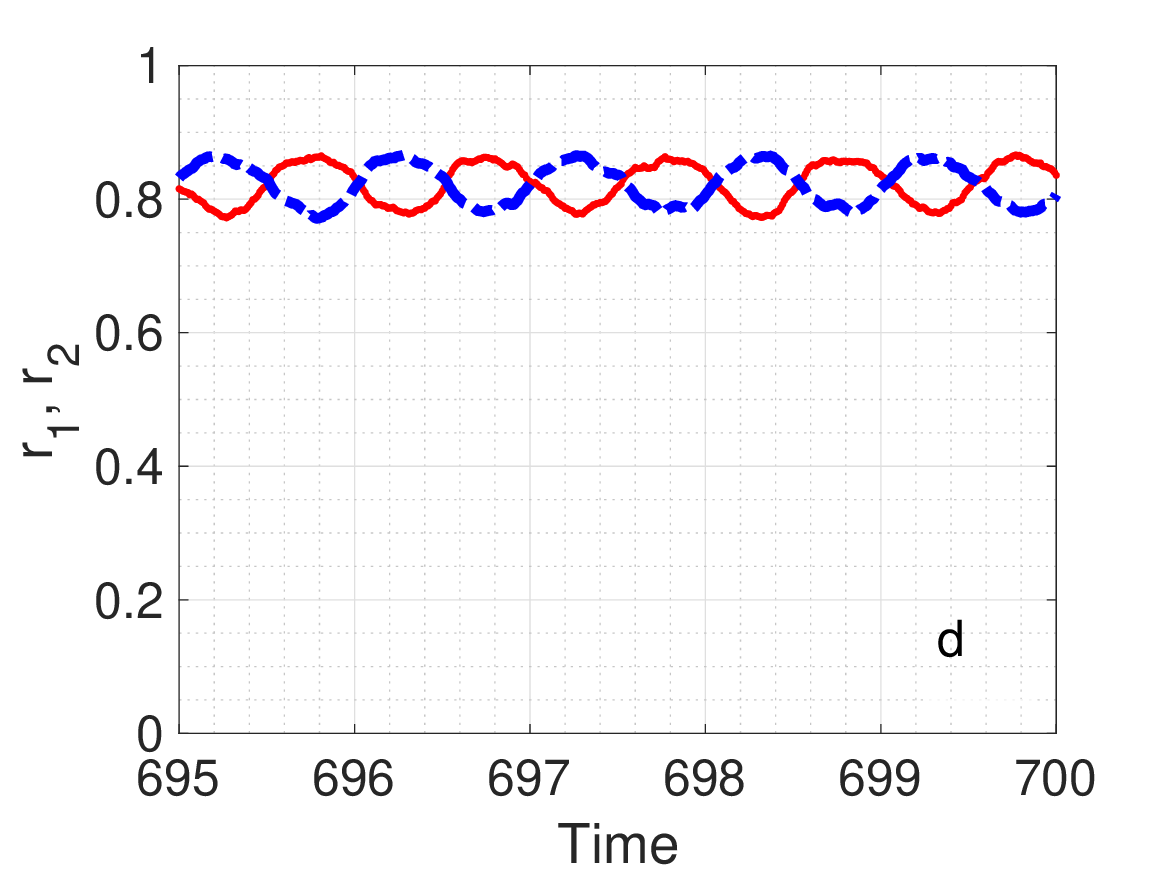}\\
  %{Simulation_timeseries_1.25.eps}\\
  \includegraphics[width=0.49\linewidth]
  {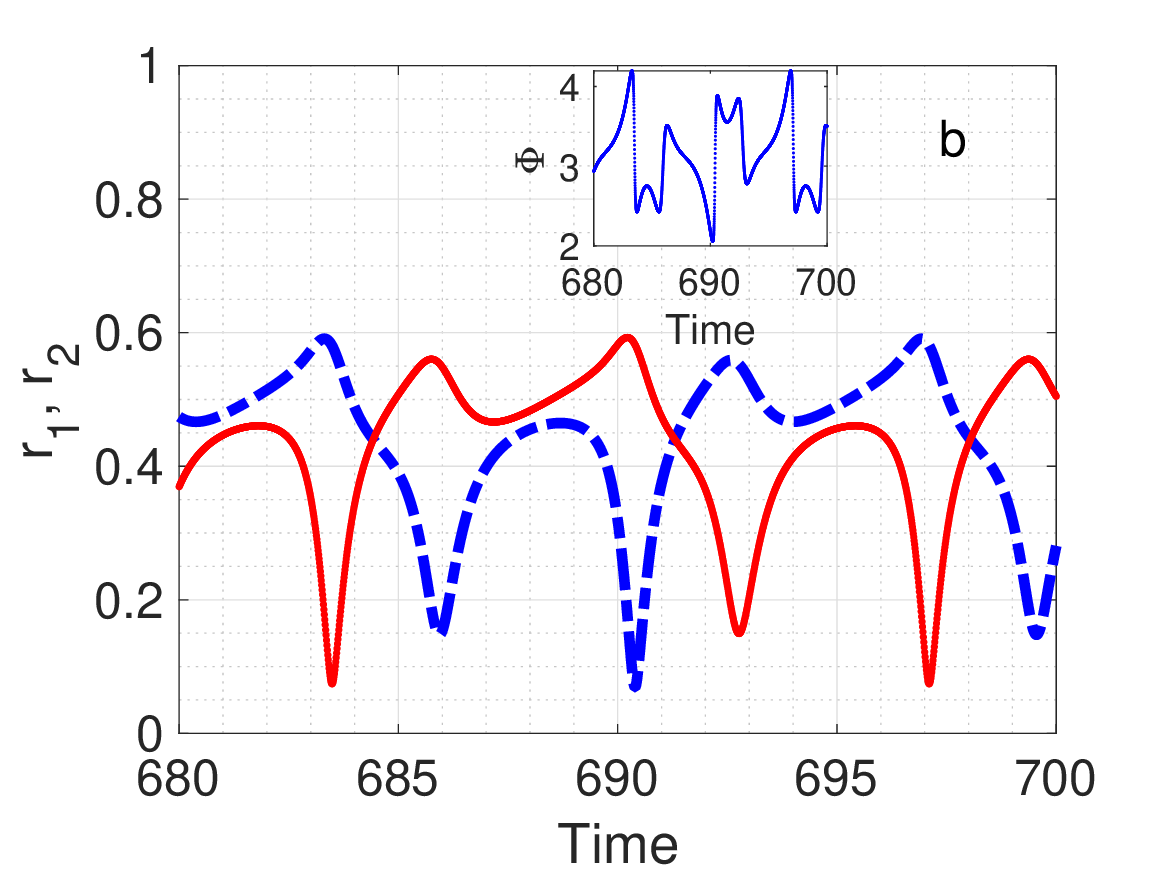}%{polar_timeseries_1.406_neg_alpha.eps}
  \includegraphics[width=0.49\linewidth]
  {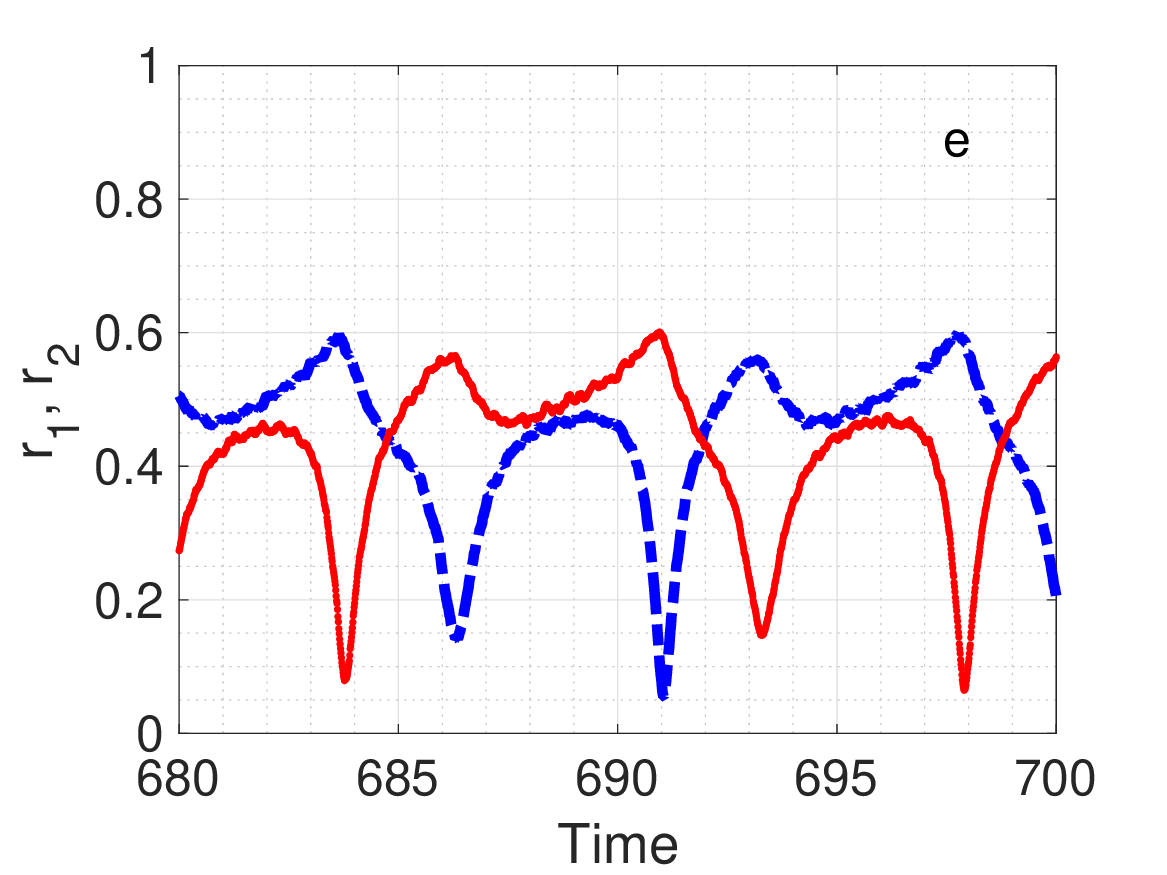}\\%{simulation_timeseries_1.406.eps}\\
 \includegraphics[width=0.49\linewidth]
 {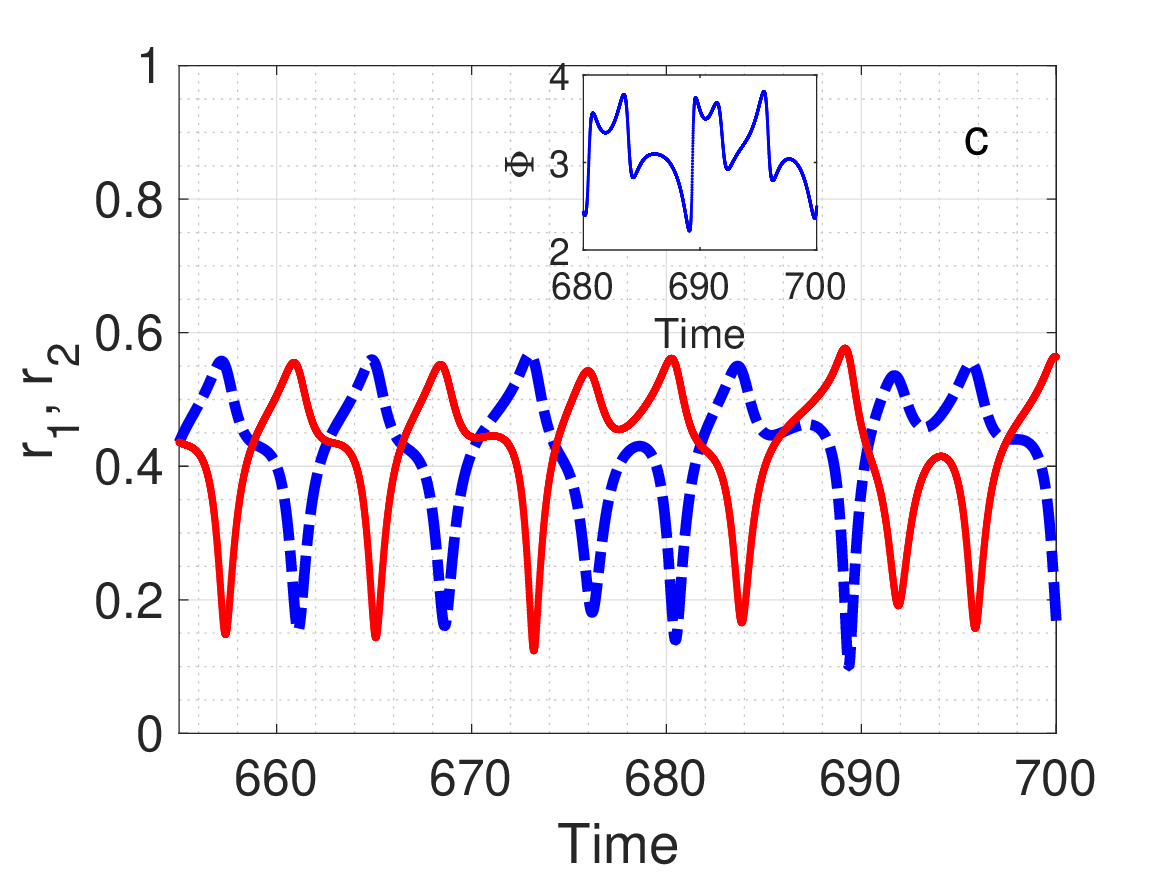}%{polar_timeseries_1.41_neg_alpha.eps}
 \includegraphics[width=0.49\linewidth]
 {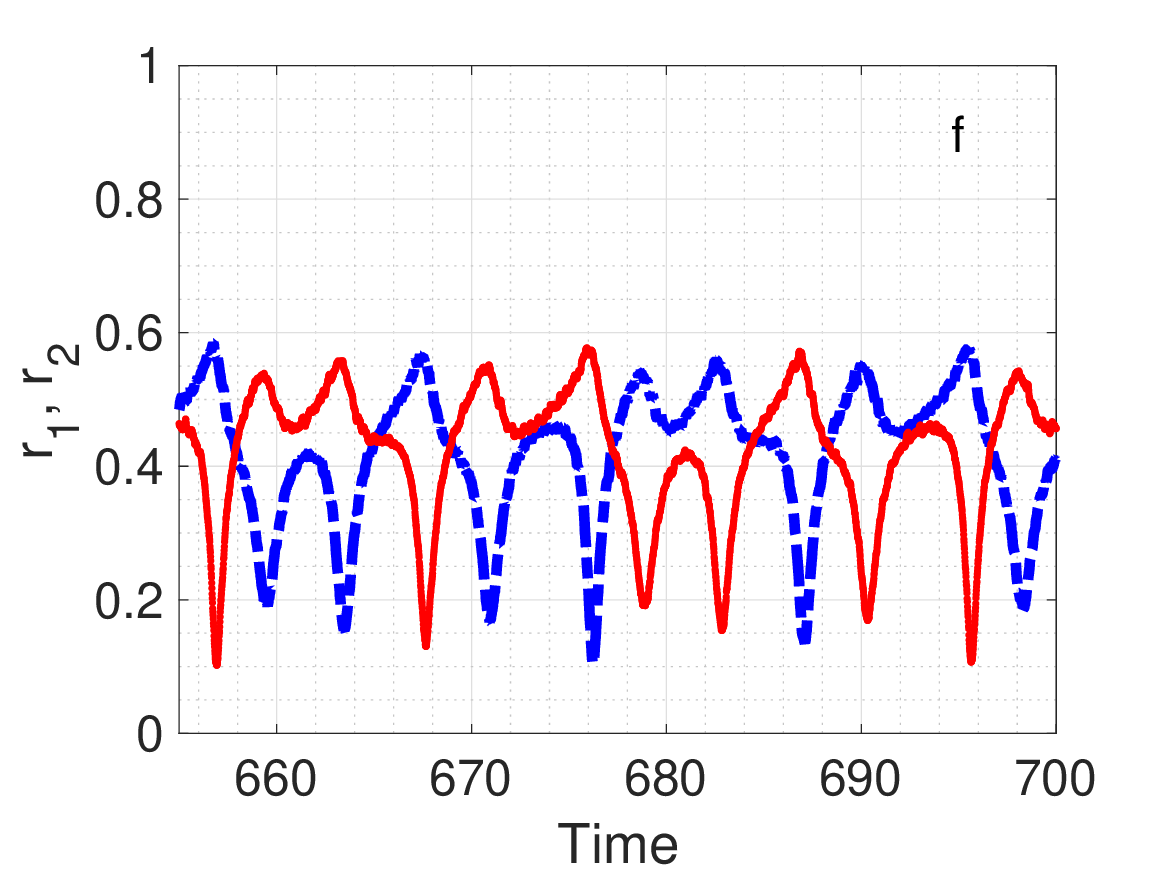}%{Simulation_timeseries_1.41.eps}
     \caption{\textit{Time series plot of the order parameters of two communities for negative $\alpha$.} Order parameters $r_1$ and $r_2$ versus time in solid red and dashed blue curves, respectively, for $K_1 = 10$, $K_2 = 10$, $\alpha = -0.5$, $\Delta = 1$, and $\gamma = 1.25$ [1(a), 1(d)], $\gamma = 1.406 $ [1(b), 1(e)], and $\gamma = 1.41$ [1(c), 1(f)]. Figs.~1(a)-(c) show order parameters found by solving Eqs.~(\ref{r1-ode})-(\ref{phi-ode}). Figs.~1(d)-(f) show order parameters found by solving Eqs.~(\ref{Kuramoto-equation}) for a system of $10000$ oscillators. The inset figures in Figs.~1(a)-(c) show the phase difference angle $\Phi$.}
     \label{fig:timerseries-theory-simulation}
\end{figure}

\begin{figure}[b]
    \centering        
    \includegraphics[width=\linewidth]
    {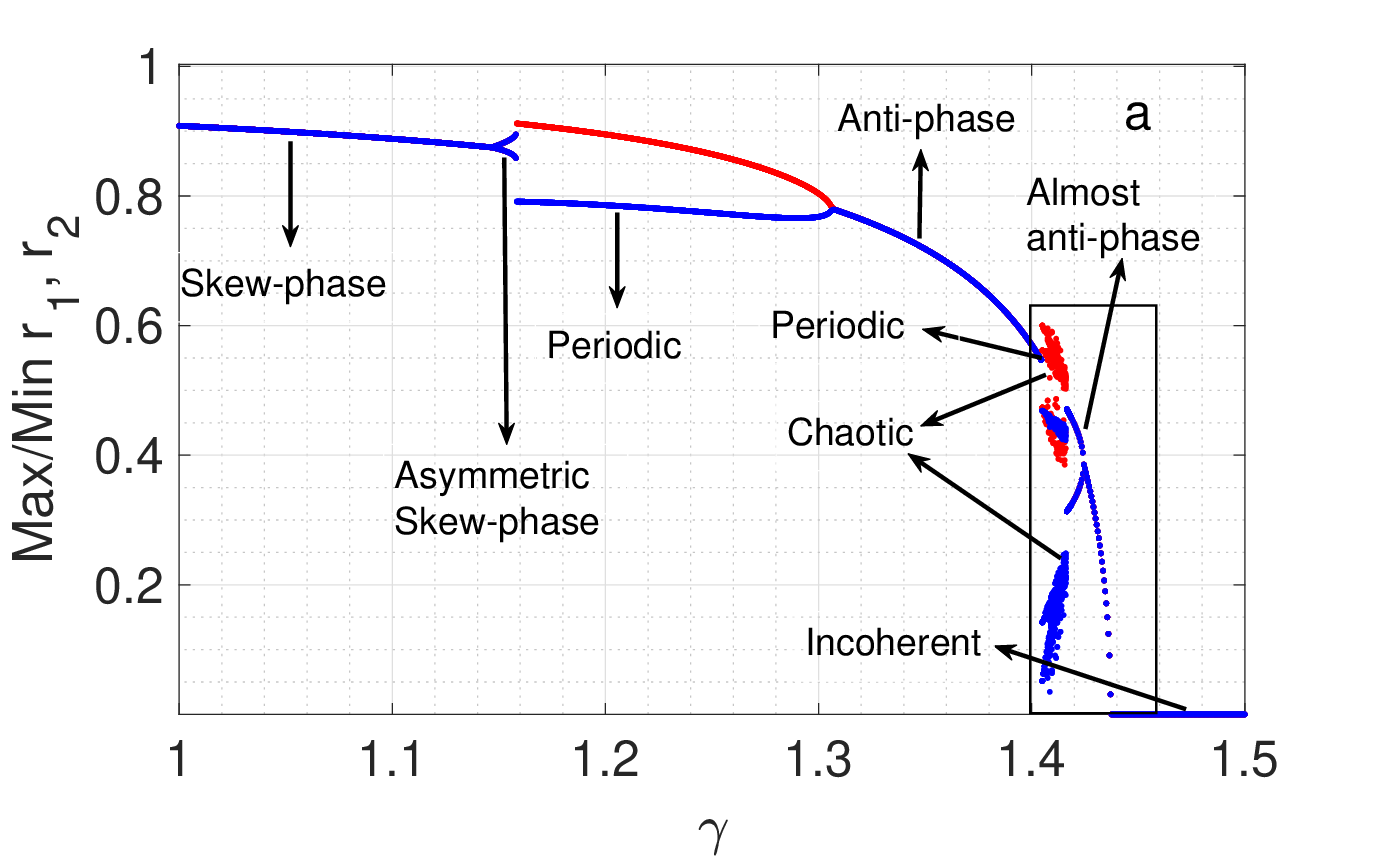}\\
     \includegraphics[width=\linewidth]
     {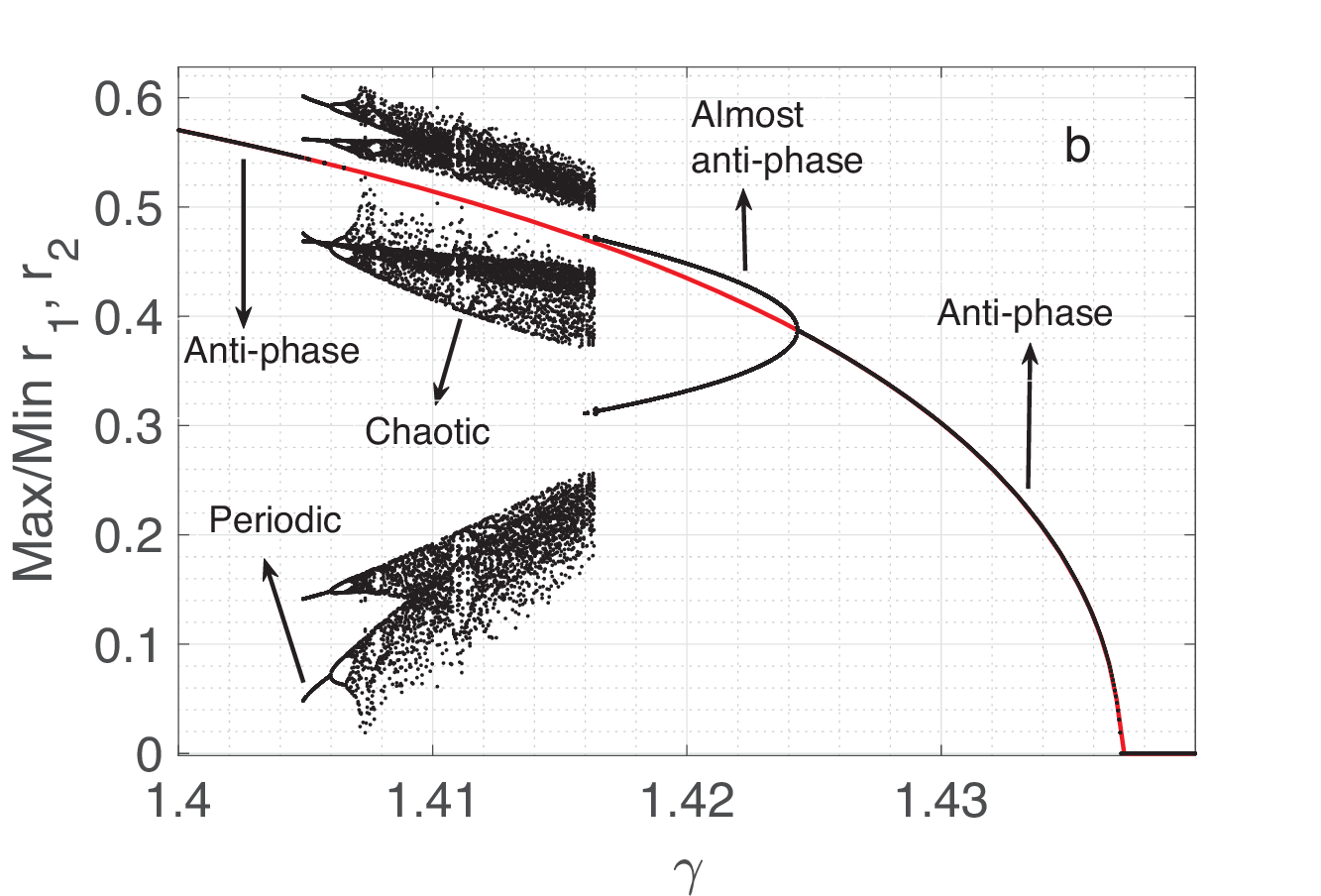}
    \caption[Bifurcation diagram for $\alpha = -0.5$]{\textit{Bifurcation diagram for $\alpha = -0.5$.} 64 local maxima (red in \ref{fig:bifurcation-chaotic-incoherent}(a) and black in \ref{fig:bifurcation-chaotic-incoherent}(b)) and 64 local minima (blue \ref{fig:bifurcation-chaotic-incoherent}(a) and black in \ref{fig:bifurcation-chaotic-incoherent}(b)) of $r_1$ and $r_2$ obtained from numerical solution of Eqs.~(\ref{r1-ode})-(\ref{phi-ode}). We set $K_1 = 10$, $K_2 = 10$, and $\Delta = 1$. (a) Bifurcation diagram in the range $\gamma \in [1,1.5]$. (b) Bifurcation diagram in the range $\gamma \in [1.4,1.44]$.}
    \label{fig:bifurcation-chaotic-incoherent}
\end{figure}

The first six possible cases (i) to (vi) are shown in Fig.~\ref{fig:steady-state-negative-alpha-complex-plane}, where the red and blue filled circles show the steady states of the two communities, i.e., the limiting values of $r_1$ (red) and $r_2 e^{i\Psi}$ (blue). When the order parameters do not converge to constant values, we plot their trajectories with blue and red lines after discarding transients. Case (vii), not shown, is similar to case (ii) but with slightly different values for $r_1$ and $r_2$.

The time series corresponding to cases (v) and (vi) are shown in Fig.~\ref{fig:timerseries-theory-simulation}, where we plot the order parameters $r_1$ (in solid red) and $r_2$ (in dashed blue) obtained by solving Eqs.~(\ref{r1-ode})-(\ref{r2-ode}) (left panels). In this and other direct numerical solutions of Eqs.~(\ref{r1-ode})-(\ref{r2-ode}) we use, unless indicated, Heun's method with time step $\Delta t = 0.001$. We choose a Lorentzian distribution with $\Delta=1$ for the natural frequencies and $K_1 = 10$, $K_2 = 10$, and $\alpha = -0.5$. 

In Fig.~\ref{fig:timerseries-theory-simulation}(a), where we choose $\gamma=1.25$, the dynamics appears to be periodic [case (v)]. We note that the oscillations of $r_1$ and $r_2$ are out of phase: when one community attains a local maximum, the other community attains a local minimum. More complex periodic behavior is seen in Fig.~\ref{fig:timerseries-theory-simulation}(b), where $\gamma = 1.406$. Fig.~\ref{fig:timerseries-theory-simulation}(c), corresponding to $\gamma = 1.41$, shows what appears to be chaotic behavior. The angle difference $\Phi$ shown in the insets is periodic when $r_1$ and $r_2$ are periodic [Fig.~\ref{fig:timerseries-theory-simulation}(a)-(b)] and chaotic when $r_1$ and $r_2$ are chaotic [Fig.~\ref{fig:timerseries-theory-simulation}(c)]. These numerical results obtained from the ODEs (\ref{r1-ode})-(\ref{phi-ode}) agree well with numerical simulations of the whole oscillator ensemble described by Eqs.~(\ref{Kuramoto-equation}), as shown in Fig.~\ref{fig:timerseries-theory-simulation} (right panels). In Fig.~\ref{fig:timerseries-theory-simulation}(d), (e), and (f), we present simulation results of Eq.~(\ref{Kuramoto-equation}) for $\gamma = 1.25$, $1.406$, and $1.41$, respectively. 

A bifurcation diagram showing how different dynamics emerge as $\gamma$ is increased is presented in Fig.~\ref{fig:bifurcation-chaotic-incoherent}(a). To produce this bifurcation diagram, we solve Eqs.~(\ref{r1-ode})-(\ref{phi-ode}) numerically for $\gamma = 1$ and, after discarding transients,  we collect $64$ local maxima and $64$ local minima for $r_1$ and $r_2$. Then we increase $\gamma$ by $5\times10^{-4}$  and repeat the procedure, adding a small random perturbation of order $10^{-2}$ to the last values of $r_1, r_2, \text{ and } \Phi$ for the previous $\gamma$ and using these perturbed values as initial conditions for the new $\gamma$, continuing until we reach $\gamma = 1.5$. The red dots show the local maxima for both communities and the blue dots show the local minima for both communities (when these are equal, only the blue dots are visible).

The bifurcation diagram is extremely rich, with bifurcations between all the seven states described above. For $\gamma$ less than approximately $1.147$, we observe the skew-phase synchronized state [case (ii)]. At approximately $\gamma = 1.147$, the skew-phase synchronized state bifurcates into the asymmetric skew-phase state [case (vii)], where $r_1$ and $r_2$ are close to, but not equal to each other. 
For $\gamma$ approximately in the interval $[1.1582, 1.306]$, the system exhibits periodic  states [case (v)]. These periodic states are followed by the anti-phase synchronized state [case (iii)] until $\gamma \approx 1.4055$. As $\gamma$ increases further, we see periodic behavior [case (v)] and a sequence of period-doubling bifurcations leading to chaotic behavior [case (vi)]. At approximately $\gamma = 1.4163$, the system suddenly falls into the almost anti-phase state [case (iv)]. Then, at $\gamma \approx 1.4243$ the two different values of $r_1$ and $r_2$ coalesce and $\Phi$ becomes $\pi$, and the system enters the anti-phase synchronized state [case (iii)]. Finally, at $\gamma \approx 1.4365$, the system reaches the incoherent state [case (i)]
\begin{figure}[b]   
    \includegraphics[width=.95\linewidth]{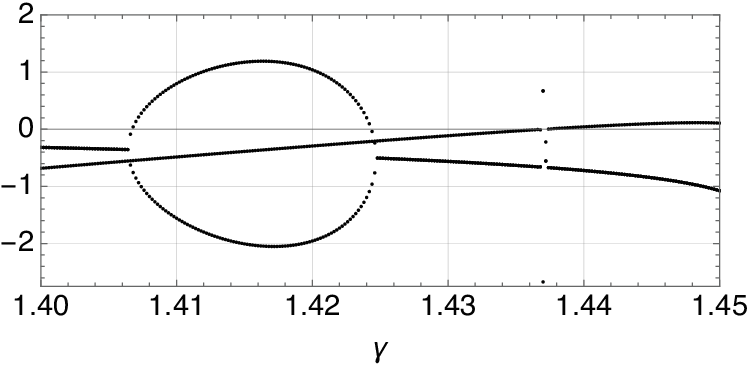}
    \caption{\textit{Real part of the eigenvalues of the Jacobian for the anti-phase solution as a function of $\gamma$.} There is an eigenvalue with positive real part, indicating instability, in the approximate region $\gamma \in [1.4055,1.4243]$ and for $\gamma > \gamma_c$. (Note that there is numerical instability when the eigenvalues cross the real axis at $\gamma = \gamma_c$.)}
    \label{fig:eigenvalues}
\end{figure}
\begin{figure}[t]
\centering
\includegraphics[width=0.85\linewidth]{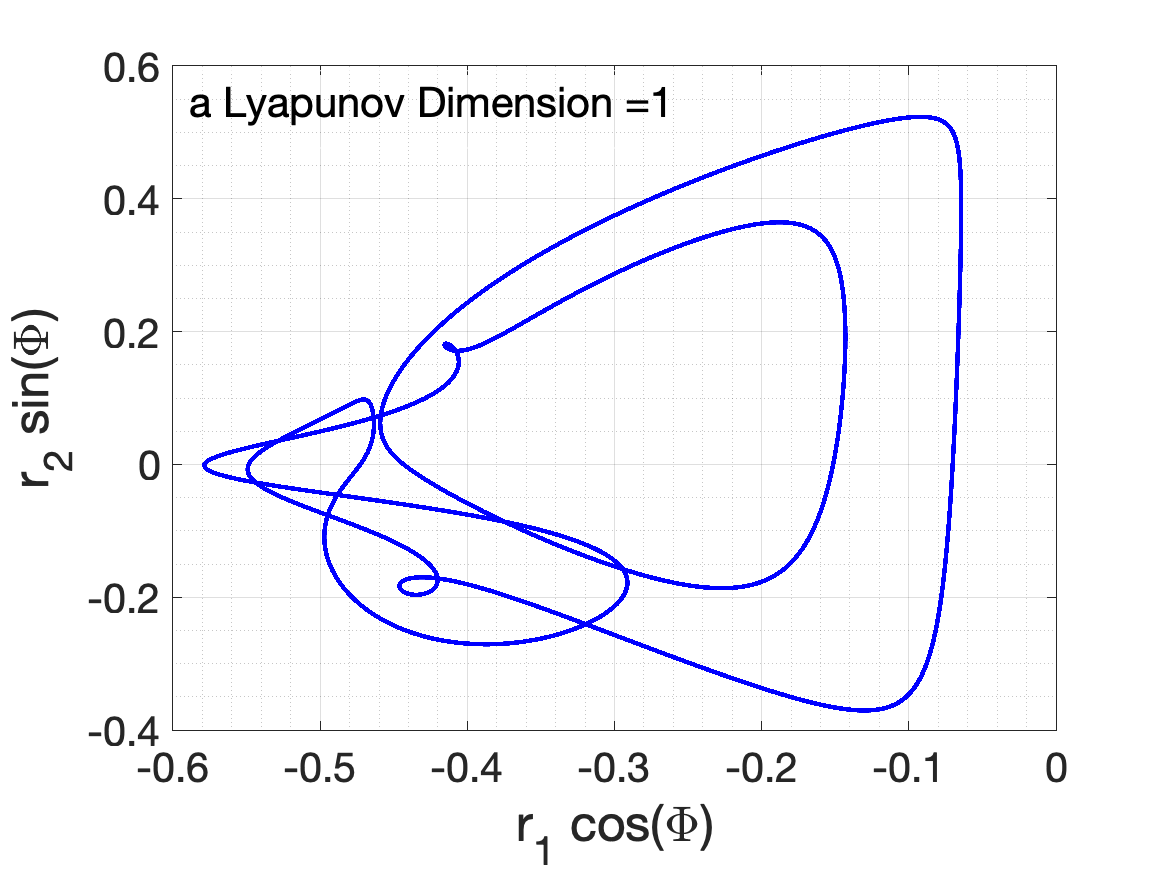}\\
\includegraphics[width=0.85\linewidth]{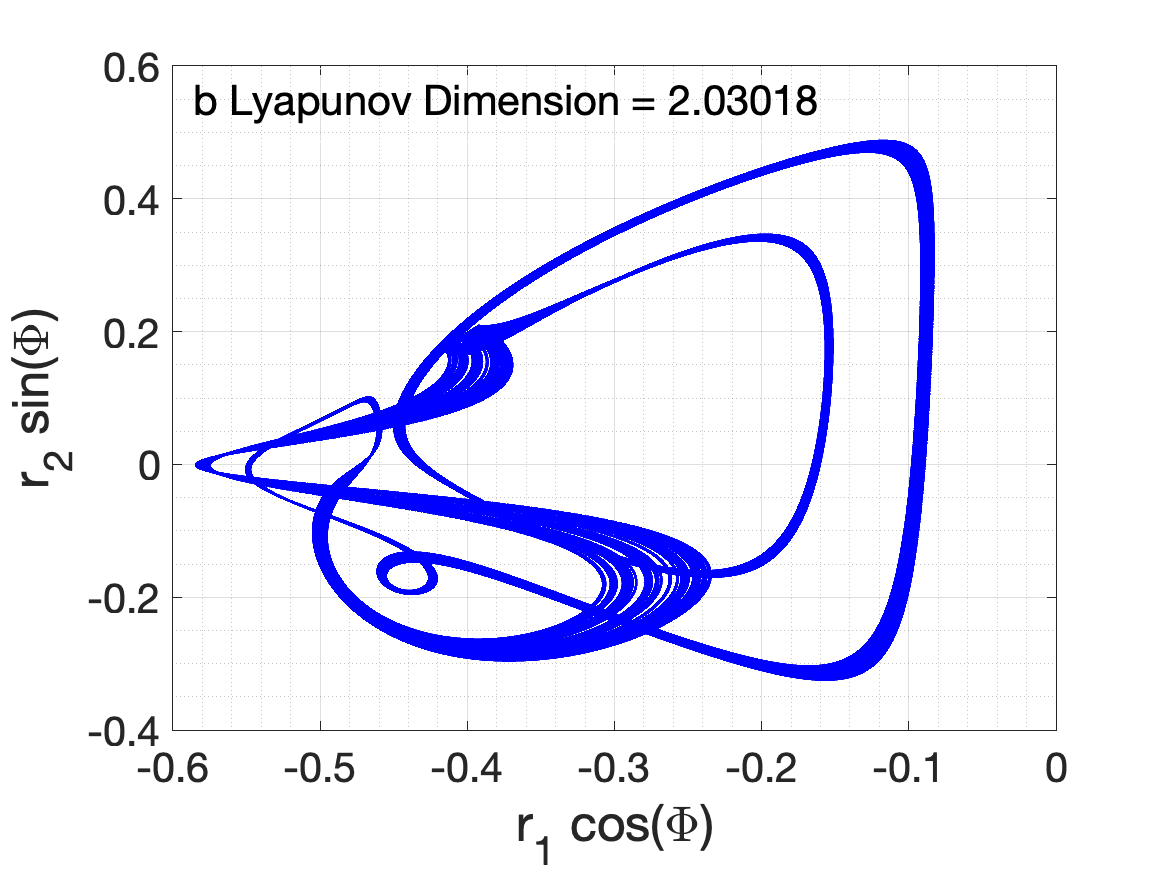}\\
\includegraphics[width=0.85\linewidth]{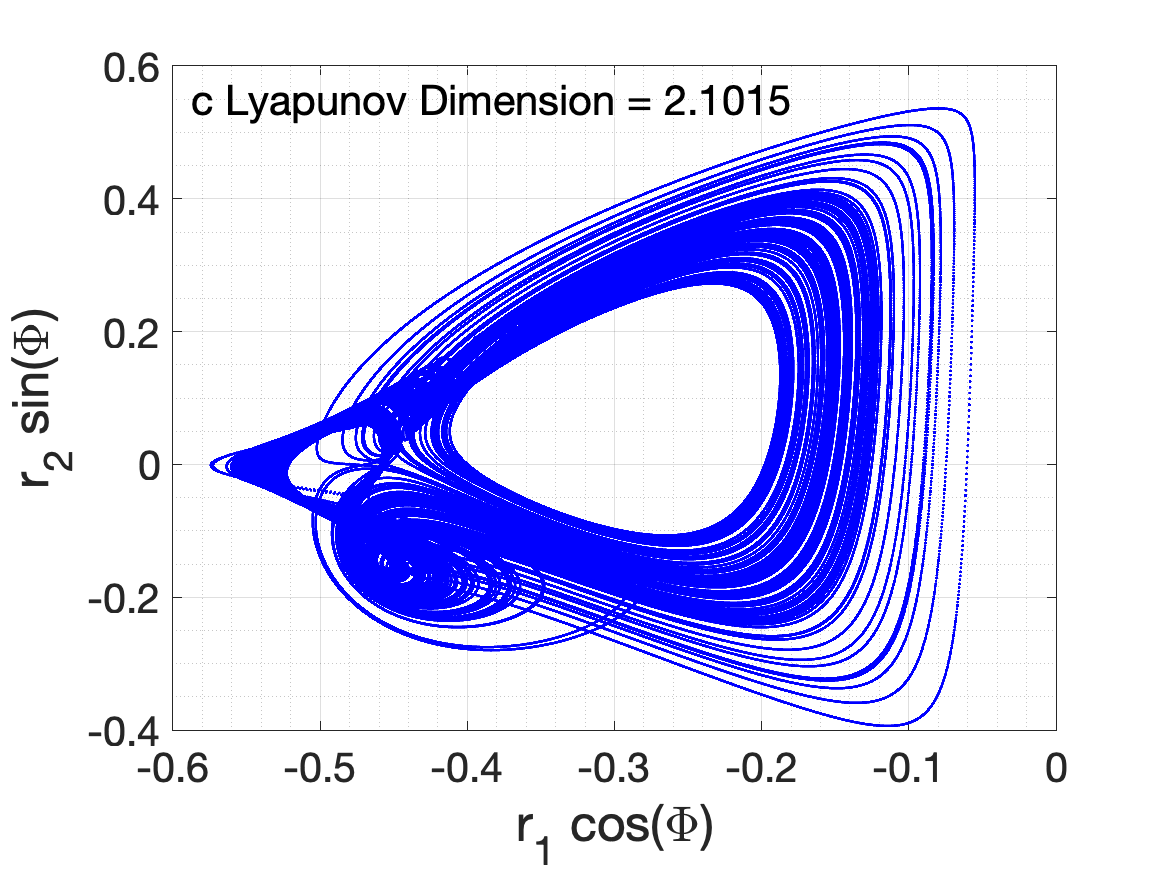}
\caption{\textit{Phase attractors of Eqs.~(\ref{r1-ode})-(\ref{phi-ode}) for $\gamma = 1.406$ (a), $\gamma=1.40685$ 3(b), and $\gamma = 1.41$ (c)}. The attractor for $\gamma = 1.406$ looks periodic, while the attractors for $\gamma=1.40685 \text{ and }1.41$ look chaotic. We set $K_1 = 10$, $K_2 = 10$, $\alpha = -0.5$, and $\Delta = 1$.}
\label{fig:attractors}
\end{figure} 

Now we discuss in more detail the bifurcations between the anti-phase synchronized, the periodic, the chaotic, the almost anti-phase, and the incoherent states observed in the boxed region of Fig.~\ref{fig:bifurcation-chaotic-incoherent}(a), which is expanded in Fig.~\ref{fig:bifurcation-chaotic-incoherent}(b) [in Fig.~\ref{fig:bifurcation-chaotic-incoherent}(b) we increase $\gamma$ in steps of $2\times 10^{-5}$]. 
First, we note that the value for $r_1 = r_2$ in the anti-phase synchronized state can be found analytically by setting $r_1 = r_2 = r$ and $\Psi = \pi$ in Eqs.~(\ref{r1-ode})-(\ref{r2-ode}), which gives 
  \begin{align}\label{antiphase}
        r_1^2 &= r_2^2 = \frac{1}{2} - \frac{1}{2(1 + \alpha)} \frac{K_1}{K_2} \\
        &+ \sqrt{ \left(\frac{1}{2} + \frac{1}{2(1 + \alpha)} \frac{K_1}{K_2}\right)^2 - \frac{2}{(1-\alpha^2)} \frac{\sec{\gamma}}{K_2}},\nonumber
\end{align}  
which is plotted in Fig.~\ref{fig:bifurcation-chaotic-incoherent}(b) with a red solid line. From this expression, we find by setting $r_1 = r_2 = 0$ that the anti-phase synchronized state disappears at the critical value of $\gamma$ given by 
\begin{align}
\gamma_c = \text{arcsec}\left[\frac{K_1(1- \alpha)}{2}\right].
\end{align}
As expected, the values of $r_1$, $r_2$, and $\gamma_c$ predicted by the theory agree very well with the numerical solutions of the differential equations. 

Using the analytical expression for the anti-phase synchronized solution, we can calculate its linear stability. To do this, we calculate numerically the eigenvalues of the Jacobian associated to Eqs.~(\ref{r1-ode})-(\ref{phi-ode}) evaluated at the values of $r_1$, $r_2$ given by Eq.~(\ref{antiphase}) and $\Phi = \pi$ as a function of $\gamma$. The real part of the three eigenvalues is shown in Fig.~\ref{fig:eigenvalues}. For the parameters used above, we find that the anti-phase synchronized state is linearly unstable in the region $\gamma \in [1.4068,1.4243]$, in agreement with the numerical observation that, in this interval, other solutions are observed in the bifurcation diagram. As we will see in Sec.~\ref{bistability}, a periodic attractor coexists with the anti-phase synchronized solution for $\gamma \in [1.4055,1.4068]$.

While the anti-phase synchronized solution is linearly unstable in the region $\gamma \in [1.4068,1.4243]$, one can also check that it is linearly stable in that interval if one restricts the dynamics to the manifold $r_1 = r_2$ [i.e., perturbations $(\delta r_1, \delta r_2,\delta \Phi)$ to the anti-phase synchronized solution with $\delta r_1 = \delta r_2$ decay exponentially]. Therefore, the instability of the anti-phase synchronized solution might not be apparent numerically if one produces a bifurcation diagram by simply increasing or decreasing $\gamma$ adiabatically, without adding a perturbation with $\delta r_1 \neq \delta r_2$ every time $\gamma$ is changed.

\subsubsection{Chaotic Dynamics}

Now we focus our attention on the chaotic dynamics [case (vi)] observed for $\gamma \in [1.4068,1.4163]$.
In Fig.~\ref{fig:attractors}, we show phase attractors for $\gamma = 1.406, 1.40685, \text{ and } 1.41$, generated by plotting $r_1 \cos{(\Phi)}$ on the horizontal axis and $r_2 \sin{(\Phi)}$ on the vertical axis. Consistent with Figs.~\ref{fig:timerseries-theory-simulation} and \ref{fig:bifurcation-chaotic-incoherent}, the attractor for $\gamma=1.406$ looks periodic, and the attractors for $\gamma = 1.40685 \text{ and } 1.41$ appear to be chaotic. To confirm the chaotic nature of the dynamics in these cases, we calculate the Lyapunov spectrum associated to the attractor of Eqs~(\ref{r1-ode})-(\ref{phi-ode}) via the $QR$ factorization method \cite{lyapunov-calculation1, lyapunov-calculation2}. While the largest Lyapunov exponent for the attractor for $\gamma=1.406$ [Fig.~\ref{fig:attractors}(a)] is zero, the chaotic attractors for $\gamma=1.40685 \text{ and } 1.41$ [Fig.~\ref{fig:attractors}(b) and Fig.~\ref{fig:attractors}(c), respectively] have positive largest Lyapunov exponents.

\begin{figure}[t]   
    \includegraphics[width=.95\linewidth]{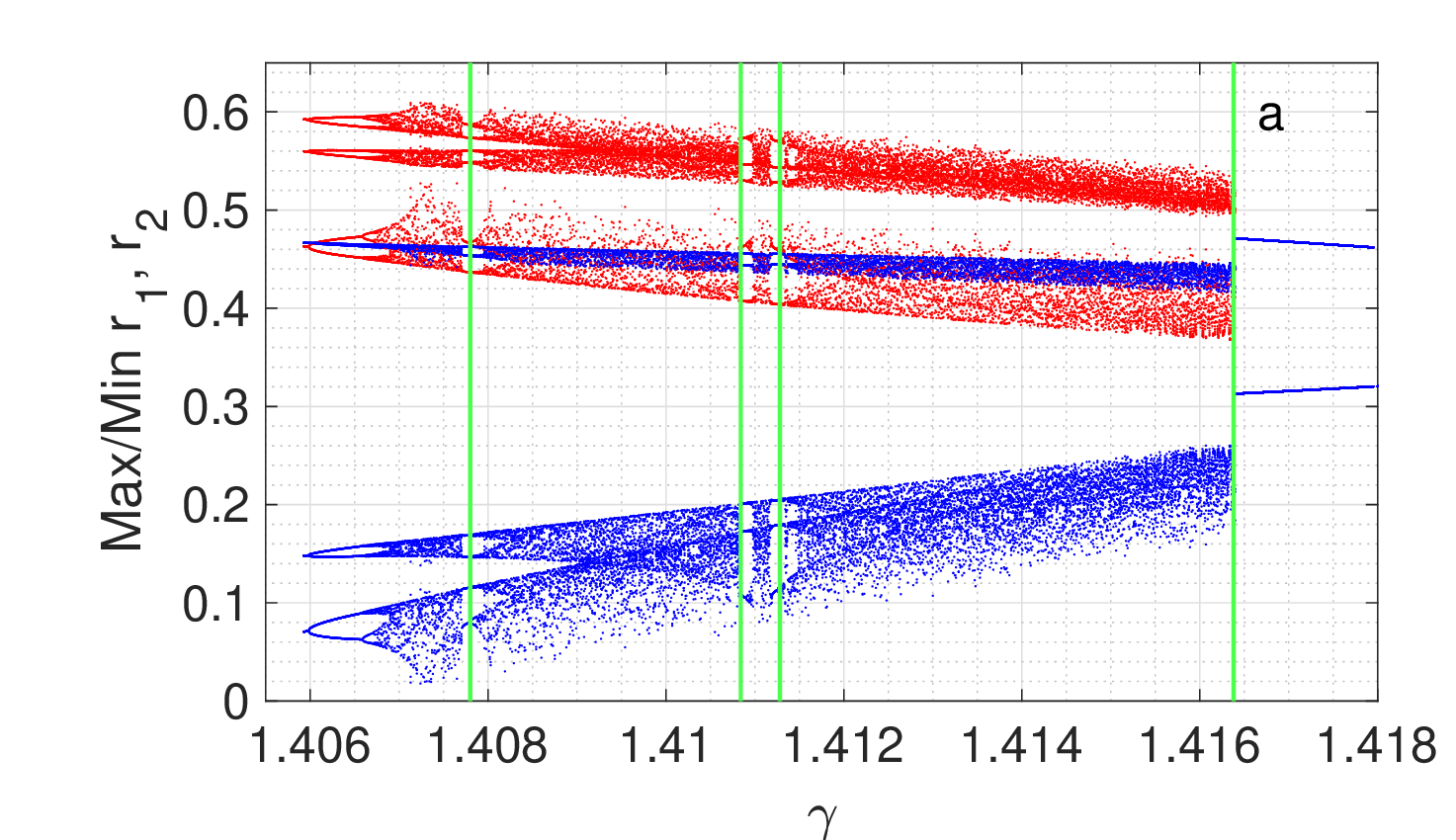}\\
    \;\;\;\;\;\;\;\includegraphics[width=.95\linewidth]{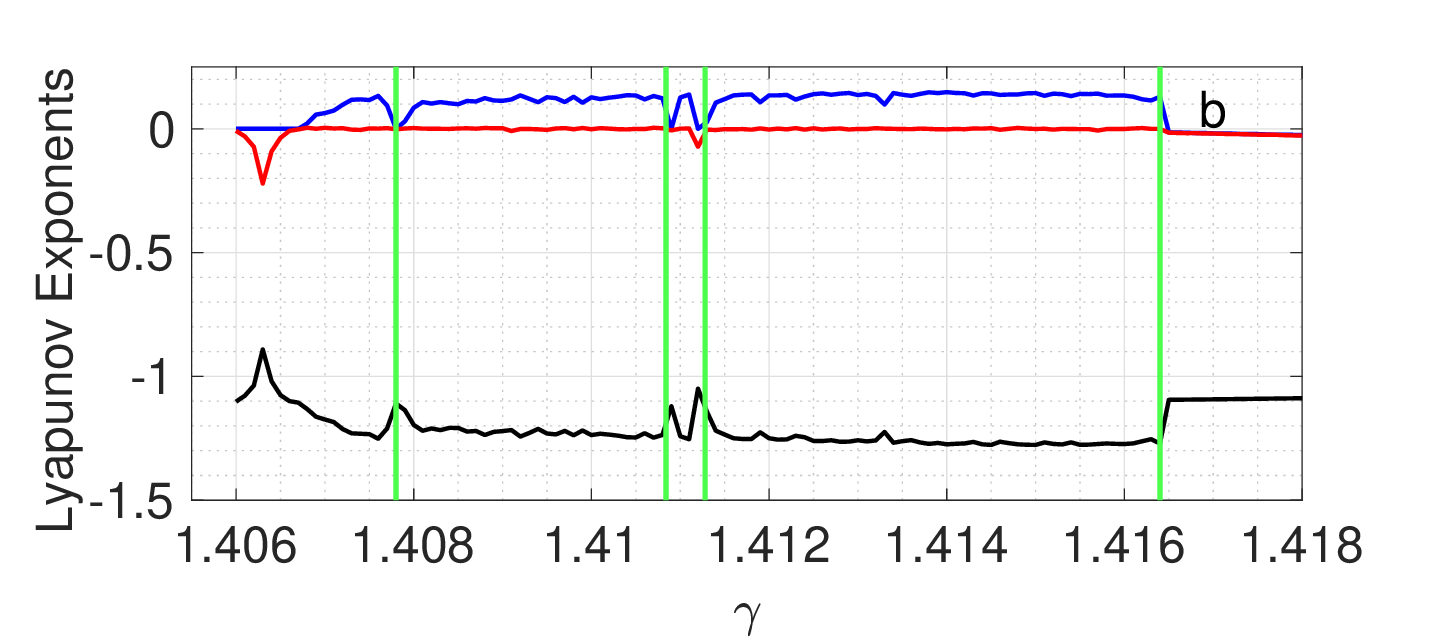}\\
    \includegraphics[width=0.95\linewidth]{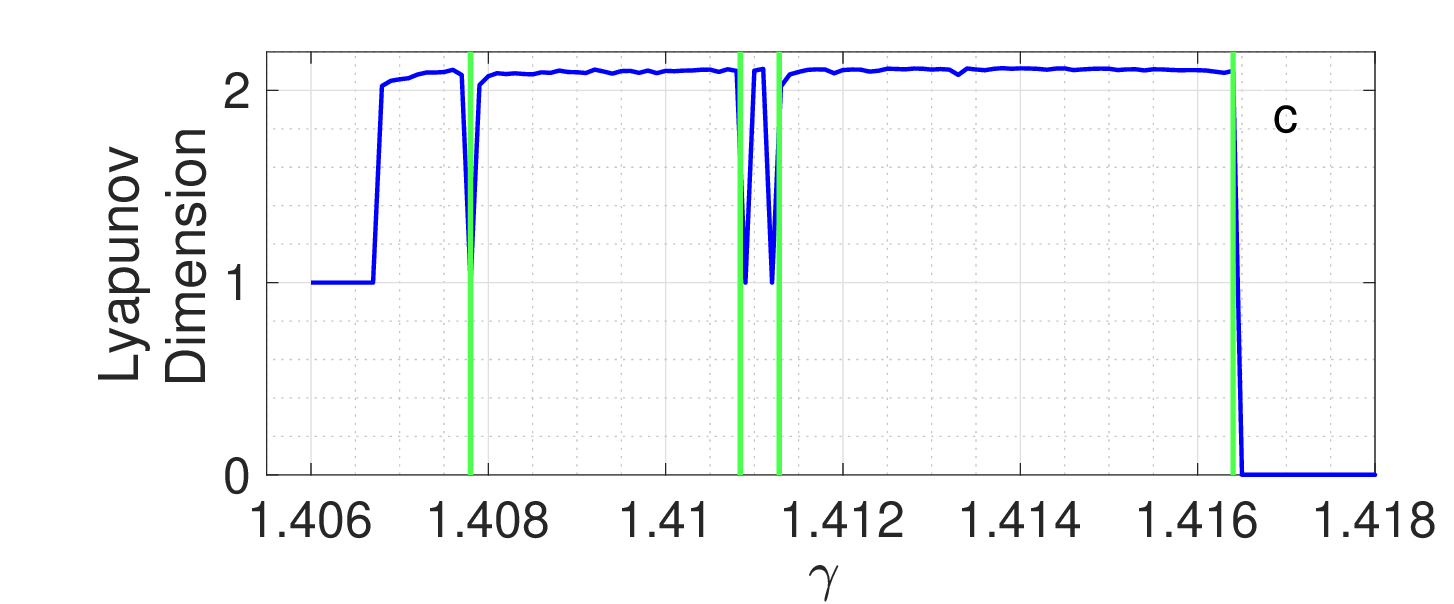}
    \caption{\textit{Bifurcation diagram, Lyapunov exponents, and the Lyapunov dimension for $\alpha = -0.5$.} 64 Local maxima and 64 local minima of $r_1$ and $r_2$ obtained from Eq.~(\ref{r1-ode})-(\ref{phi-ode}) are plotted in red and blue, respectively in panel (a). Panels (b) and (c) show the Lyapunov exponents and the Lyapunov dimension, respectively, of the attractors associated to Eqs.~(\ref{r1-ode})-(\ref{phi-ode}). We observe periodic behavior for $\gamma < 1.4068$, chaotic behavior for $\gamma \in [1.4068, 1.4163]$, and the almost anti-phase synchronized state for $\gamma \in [1.4163, 1.418]$. The green solid lines indicate the occurrence of some periodic windows within the chaotic regime.  We set $K_1 = 10$, $K_2 = 10$, and $\Delta = 1$.}
    \label{fig:bifurcation}
\end{figure}

A more detailed analysis is shown in Fig.~\ref{fig:bifurcation}, which shows the bifurcation diagram (a), the Lyapunov exponents (b), and the Lyapunov dimension \cite{lyapunov-dimension1, hilborn_lyapunov_2017} (c) in the range $1.4055 \leq \gamma \leq 1.418$. These results confirm the existence of a robust parameter range with chaotic dynamics.

\subsubsection{Bistability}\label{bistability}

\begin{figure}[b]
    \centering
    \includegraphics[width=0.8\linewidth]
    {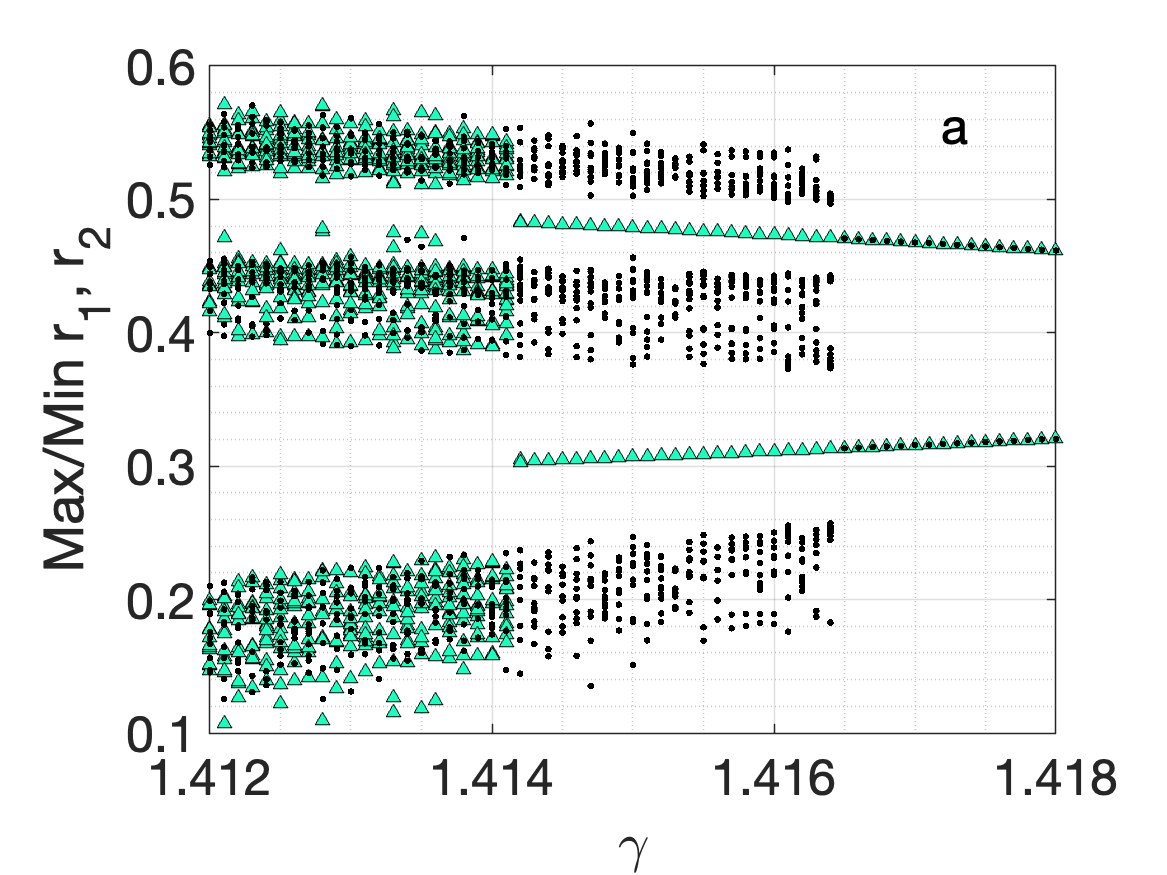}
    \includegraphics[width=0.8\linewidth]
    {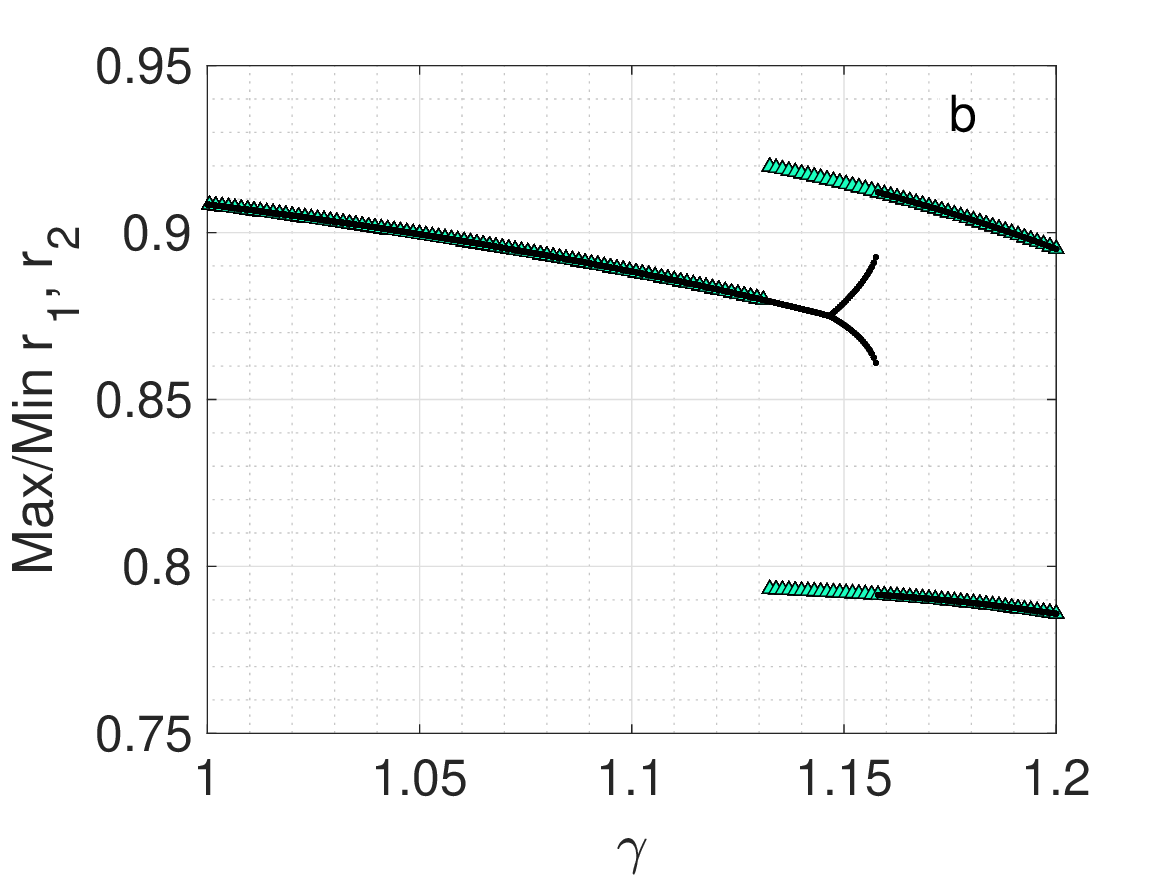}
    \caption{\textit{Bistability diagrams in the periodic and the chaotic regime.} {The black markers show the local maxima and minima when we simulate Eqs.~(\ref{r1-ode})-(\ref{phi-ode})} when increasing $\gamma$. The green filled triangles show the local maxima and minima when decreasing $\gamma$. The upper panel shows bistability in the chaotic regime whereas the lower panel shows bistability in the periodic regime.}
    \label{fig:bistability}
\end{figure}

There are various regions of parameter space in which different solutions coexist. This is not apparent in Fig.~\ref{fig:bifurcation-chaotic-incoherent}, which was produced by adiabatically increasing $\gamma$, but can be seen by comparing this bifurcation diagram with another produced by decreasing $\gamma$, or alternatively by choosing initial conditions at random for a given set of parameters.

In Figure~\ref{fig:bistability}(a) we plot the local maxima and minima of $r_1$ and $r_2$ when increasing $\gamma$ (black markers) and when decreasing $\gamma$ (green filled triangles), obtained from a  numerical solution of Eqs.~(\ref{r1-ode})-(\ref{phi-ode}). We see that for $\gamma \in [1.4144,1.4163]$ the system can be either in the chaotic state or in the almost anti-phase state, depending on the initial conditions.

Similarly, Fig.~\ref{fig:bistability}(b) shows the same quantities for $\gamma$ in the interval $[1, 1.2]$. As $\gamma$ is increased, the skew-phase solution bifurcates into the asymmetric skew-phase solution, which in turn leads to the periodic solution. As $\gamma$ is decreased, however, the periodic solution persists down to almost $\gamma \approx 1.1325$, where the system falls into the skew-phase solution.

\begin{figure}[t]
    \centering
    \includegraphics[width=\linewidth]{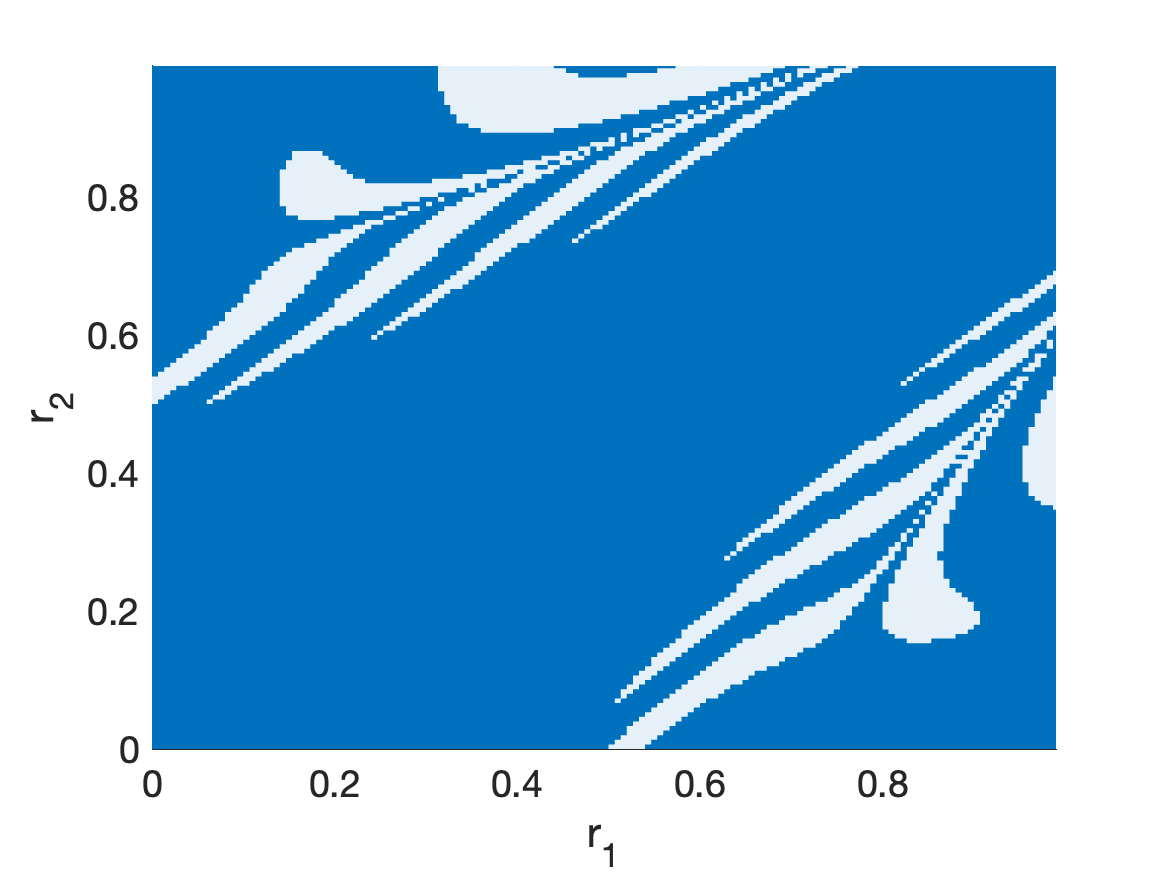}
    %{basin_phase=1.415_new.pdf}
    \caption{\textit{Basins of attraction of a fixed point (in white) and a chaotic solution (in blue).} 150 different initial conditions of $r_1$ and $r_2$ are used to generate the basins of attraction. We set $K_1 = 10$, $K_2 = 10$, $\alpha = -0.5$, $\gamma = 1.416$, $\Delta = 1$, and $\Phi(0) = \pi/6$.} 
    \label{fig:basin-of-attraction}
\end{figure}

Figure \ref{fig:basin-of-attraction} shows the basins of attraction of the chaotic and fixed point solutions for $\gamma = 1.416$. For a given $r_1$, $r_2$, we fix $\Phi = \pi/6$ and simulate Eqs.~(\ref{r1-ode})-(\ref{phi-ode}) for 700 time units. After discarding transient effects, we determine whether the solution is a fixed point or chaotic by examining the difference between maximum and minimum values of the $r_1$ and $r_2$ time series. If the difference is larger than 0.02 (smaller than 0.02), we classify the solution as chaotic (a fixed point) and indicate it with a blue (white) pixel. As we can see, while most initial conditions lead to the chaotic state, the fixed point solution is attained for some initial conditions. We also note that the basin is not symmetric across the $r_1 = r_2$ axis, because the original equations are symmetric under the transformation $r_1 \leftrightarrow r_2$, $\Phi \leftrightarrow -\Phi$, but $\Phi$ is maintained constant at $\Phi = \pi/6$.
%%%%%%%%%%%%%%%%%%%%%%%%

Before moving on, we note that the results shown above are all for a particular choice of parameters ($K_1=10$, $K_2=10$, and $\alpha = -0.5$) as our focal interest is understanding the effects of phase delays. 

%%%%%%%%%%%%%%%%%%%%%%%%%%%%
\subsection{Attractive coupling, $0 < \alpha < 1$}\label{positivecoupling}

In the previous Section, we found that the combination of higher-order interactions, community structure, and phase lags can produce very rich dynamics, including chaotic dynamics. However, we used a negative value of $\alpha$. Since negative values of $\alpha$ correspond to negative pairwise coupling across communities, which typically enrich the dynamics, one might wonder if the chaotic dynamics can also be found for positive values of $\alpha$. In this Section we briefly address this issue by showing that chaotic dynamics are also possible for positive values of $\alpha$.

\begin{figure}
    \centering \includegraphics[width=80mm]{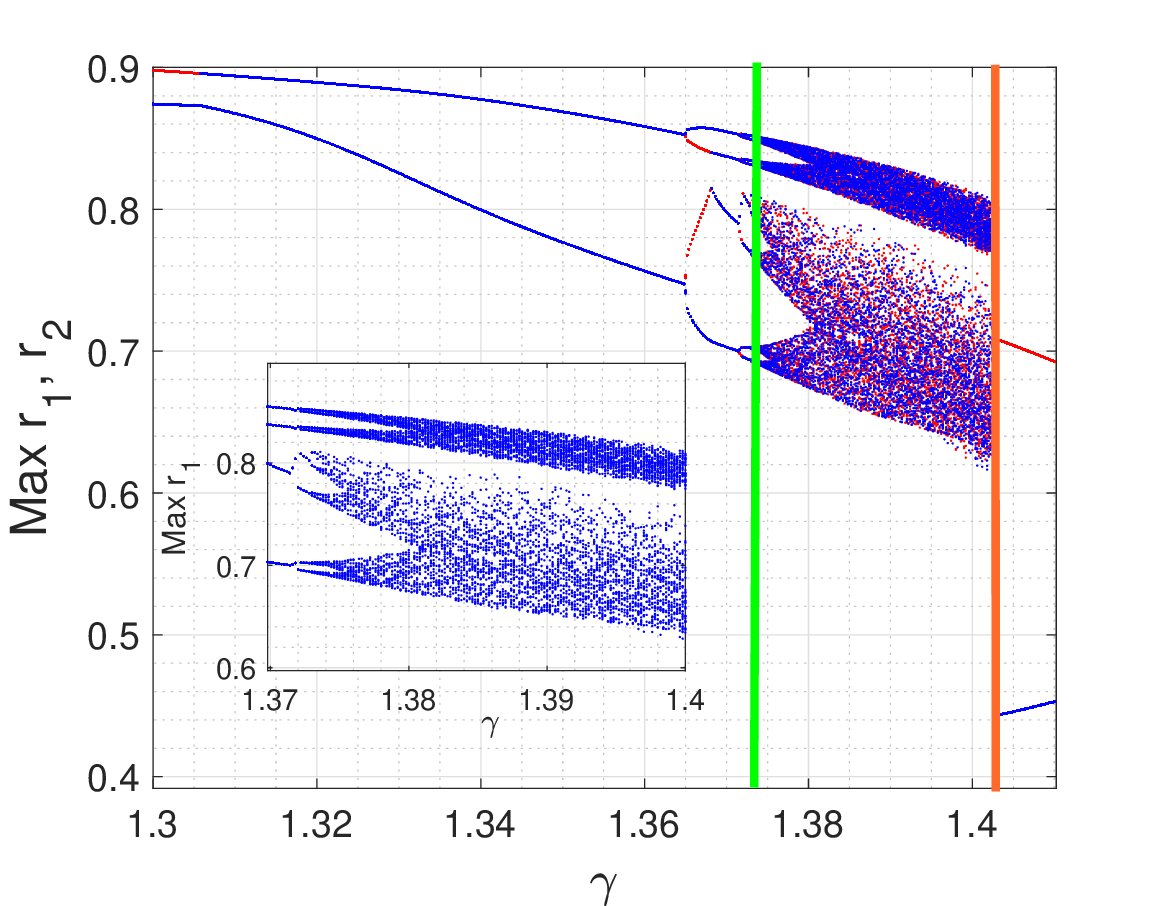}
    \includegraphics[width=80mm]{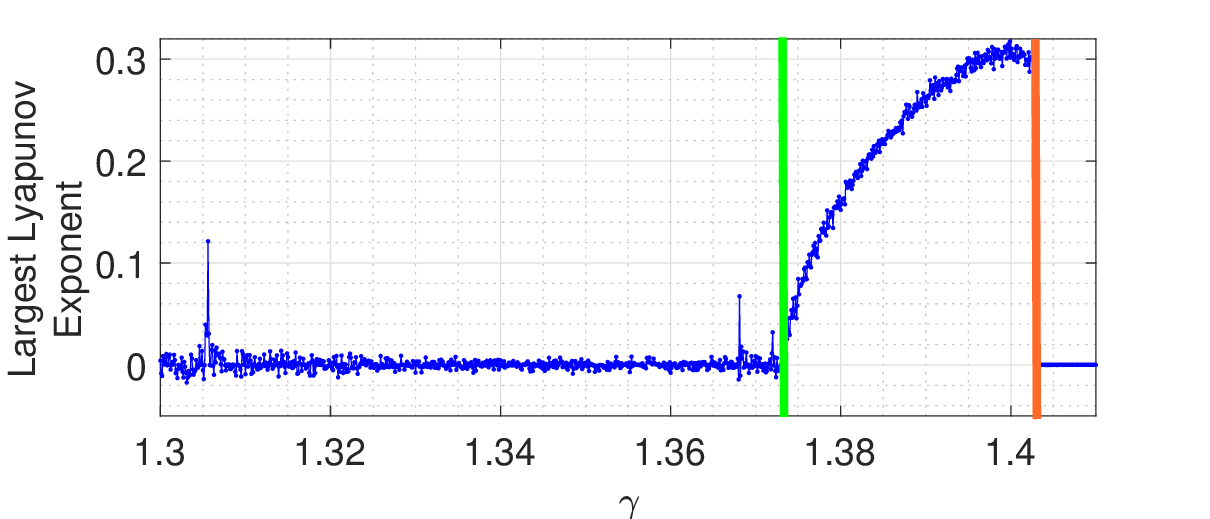}
    \caption{\textit{Bifurcation diagram and the largest Lyapunov exponent for $\alpha = 0.5$.} In panel (a), 64 local maxima of $r_1$ and $r_2$ obtained from Eqs.~(\ref{r1-ode})–(\ref{phi-ode}) are plotted in red and blue, respectively, for $K_1 = 10$, $K_2 = 10$, and $\Delta = 1$. The inset in panel~(a) zooms into the chaotic regime. In panel~(b), the largest Lyapunov exponent is plotted using a blue curve. The vertical solid green line on the left and the vertical solid orange line on the right mark the onset and the end of chaotic regime, respectively.}
    \label{fig:bifurcation-positive-alpha}
\end{figure}

%As in the bifurcation diagram for $\alpha = -0.5$ (Fig.~\ref{fig:bifurcation}), 
In Fig.~\ref{fig:bifurcation-positive-alpha}(a) we show a bifurcation diagram obtained by plotting 64 local maxima of $r_1$ (red) and $r_2$ (blue) obtained from numerical solution of Eqs.~(\ref{r1-ode})-(\ref{phi-ode}) using Heun's method with time increment $\Delta t = 0.001$ for $\alpha=0.5$, $K_1=10, K_2 = 10$, and $\Delta = 1$. We collect the maxima and minima after evolving Eqs.~(\ref{r1-ode})-(\ref{phi-ode}) for 700 time units so that transient effects disappear. The system exhibits periodic behavior, including a period-doubling cascade, for $\gamma < 1.373$ (vertical green line), after which the system enters a chaotic regime. Starting at $\gamma \approx 1.4035$ (vertical orange line), the two communities attain the almost in-phase state. In Fig.~\ref{fig:bifurcation-positive-alpha}(b), we plot the largest Lyapunov exponent of Eqs.~(\ref{r1-ode})-(\ref{phi-ode}) for $\alpha=0.5$ to verify the periodic and chaotic nature of the dynamics. In agreement with the discussion above, the positive largest Lyapunov exponent shown in Fig.~\ref{fig:bifurcation-positive-alpha}(b) suggests chaotic behavior for $\gamma \in [1.373, 1.4035]$.

%%%%%%%%%%%%%%%%%%%%%%%%%%%%%%%%%%%%%%%%%%%
\begin{figure}
    \centering   \includegraphics[width=0.475\linewidth]{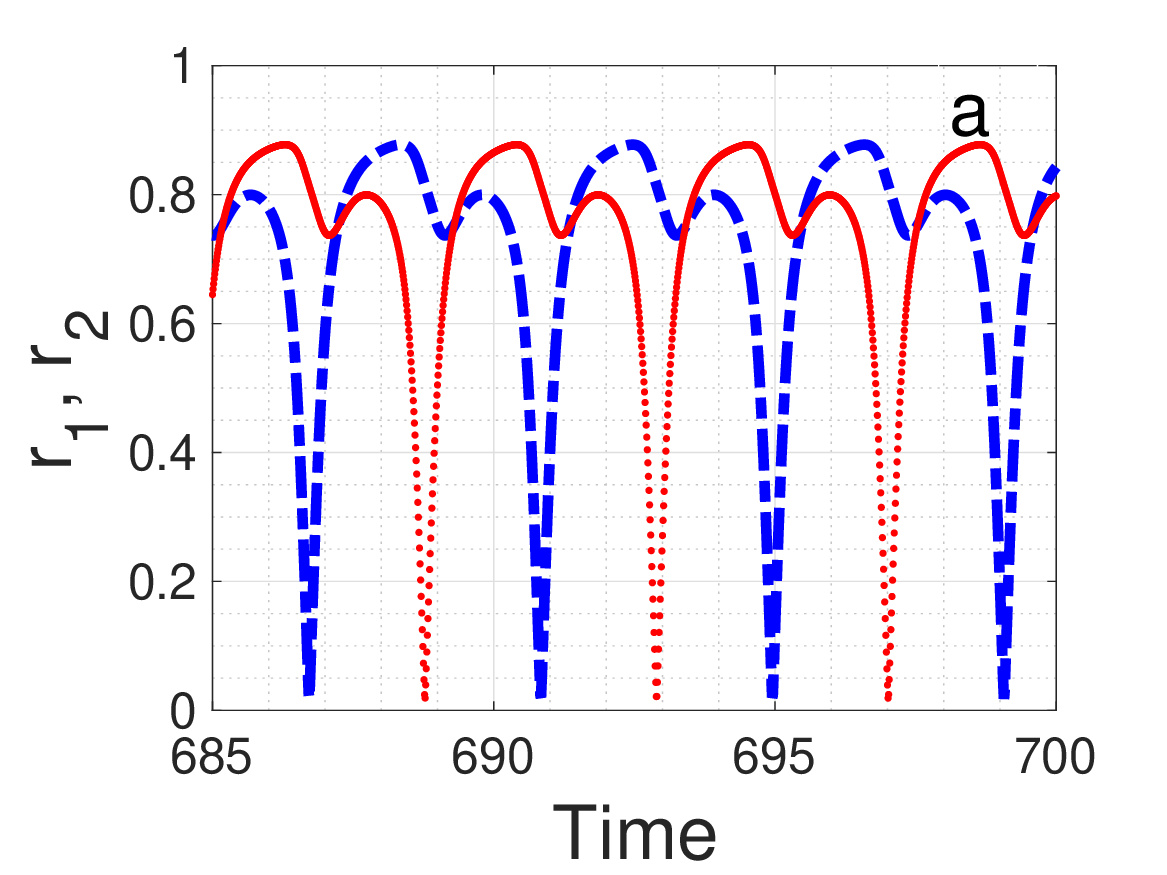}   \includegraphics[width=0.475\linewidth]{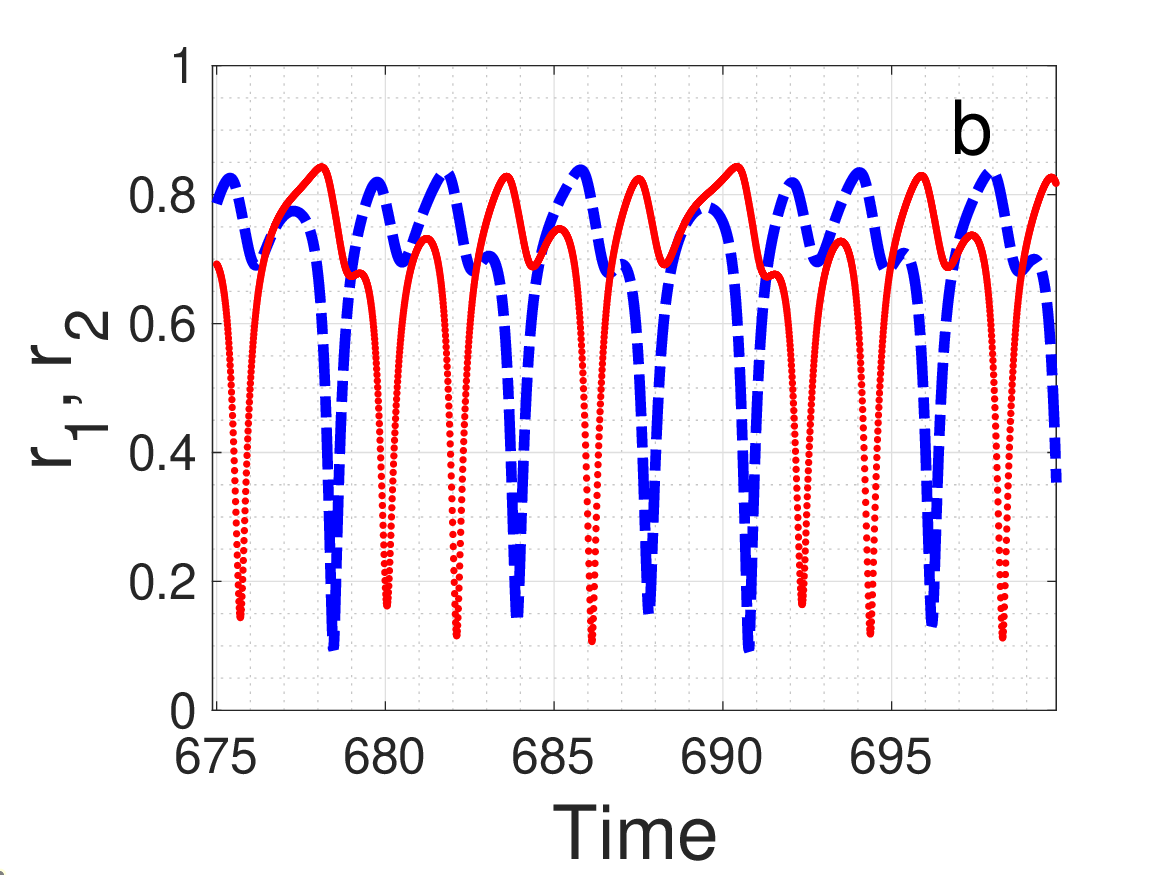}%{polar_timeseries_positive_1.34.pdf}    \includegraphics[width=0.475\linewidth]{polar_timeseries_positive_1.38.pdf}
    \caption{\textit{Time series plot of the order parameters $r_1$ (red solid curve) and $r_2$ (blue dashed curve) for $\gamma = 1.34$ and $\gamma = 1.38$.} We set $K_1 = 10$, $K_2 = 10$, $\alpha = 0.5$, and $\Delta = 1$. The system exhibits periodic synchronization behavior for $\gamma=1.34$ (a) and chaotic synchronization behavior for $\gamma=1.38$ (b).}    \label{fig:polar_timeseries_positive_alpha}
\end{figure}
%%%%%%%%%%%%%%%%%%%%%%%%%%%%%%%%%%%%
\section{Discussion \label{Discussion}}
In this work, we have studied synchronization of phase oscillators with higher-order interactions, community structure, and phase lags. When the phase lag is small, higher-order interactions and community structure result in different steady synchronized states, as found in Ref.~\cite{Multistability-communities}. For stronger phase lag values, the system exhibits a wide range of dynamics including different levels of synchrony, periodic synchronization, or chaotic synchronization. We illustrated these dynamics for two cases: $\alpha = -0.5$ and $\alpha=0.5$.

Negative $\alpha$ implies contrarian coupling between oscillators in different communities. For $\alpha = -0.5$, in addition to the incoherent state and the skew-phase synchronized state found in Ref.~\cite{Multistability-communities}, the oscillators attain four additional steady states: periodic order parameters, chaotic order parameters, the anti-phase synchronized state, and the almost anti-phase state, depending on the values of the phase lag.

As in the case of $\alpha = -0.5$, we again observe steady, periodic, and chaotic dynamics for $\alpha = 0.5$. Unlike for negative $\alpha$ values, both communities either attain local maxima or local minima simultaneously. Moreover, the communities are almost in-phase while reaching two different synchronized states.

This work sheds light on interesting and richer dynamics which are not present when only communities and higher-order interactions are considered in an ensemble of coupled oscillators. Moreover, in contrast to \cite{Chaos-phaselags}, we show that homogeneous phase lags are sufficient to induce chaotic behavior in coupled oscillators when there are higher-order interactions. We conjecture that different phase lag strengths in the pairwise and triadic coupling should also result in rich dynamics. 

Despite many interesting results, there are a few limitations in this work. We demonstrate our results by using specific cases of coupling strengths and $\alpha$ values. As our main interest was to explore whether homogeneous phase lags with higher-order interactions could result in chaos, we fixed the coupling strengths and $\alpha$ and only varied phase lag values. We have not systematically explored other values of $\alpha, K_1, \text{ and } K_2$. Understanding how the interplay of the community structure and coupling strengths affect such complex dynamics is one direction for future work. Additionally, we considered the synchronization dynamics in complete hypergraphs with two communities only. How these dynamics change in hypergraphs with more complex structure is yet to be explored.
%%%%%%%%%%%%%%%%%%%%%%%%%%%%%
%%%%%%%%%%%%%%%%%%%%%%%%%%%%%

\section*{acknowledgements}

JGR acknowledges support from {the National Science Foundation (grant number DMS-2205967).}

\section*{Data Availability}

The codes and the datasets analyzed in this study are available in GitHub at {https://github.com/saad1282/Chaos-in-a-phase-oscillator-system-with-higher-order-interactions-communities-and-phase-lags}.

\nocite{*}

\bibliography{apssamp}% Produces the bibliography via BibTeX.

\end{document}